\documentclass[journal=nalefd,manuscript=letter,layout=traditional]{achemso}

\graphicspath{{./figures/}}
\usepackage{amsmath,amsfonts,amsthm,amssymb}
\usepackage[dvipsnames]{xcolor}
\usepackage[pdftex,colorlinks=true,linkcolor=blue,citecolor=blue,urlcolor=blue]{hyperref}
\usepackage[capitalise]{cleveref}
\usepackage[utf8]{inputenc}
\usepackage[english]{babel}
\addto\captionsenglish{}
\usepackage[]{subfigure}

\usepackage[colorinlistoftodos, textsize=footnotesize]{todonotes}

\newcommand{\kbT}{k_\textnormal{B} T}

\title{Bridging Microscopic Dynamics and Hydraulic Permeability
	in Mechanically-Deformed Nanoporous Materials}

\author{Alexander Schlaich}
\email{alexander.schlaich@simtech.uni-stuttgart.de}
\affiliation{ Stuttgart Center for Simulation Science (SC SimTech),
University of Stuttgart, 70569 Stuttgart, Germany}
\alsoaffiliation{ Institute for Computational Physics, University of Stuttgart,
   70569 Stuttgart, Germany}
\alsoaffiliation{ Univ.\ Grenoble Alpes, CNRS, LIPhy, 38000 Grenoble, France}
\author{Matthieu Vandamme}
\affiliation{ Navier, Ecole des Ponts, Univ.\ Gustave Eiffel, CNRS,
   Marne-la-Vall\'ee, France}
   \author{Marie Plazanet}
\affiliation{ Univ.\ Grenoble Alpes, CNRS, LIPhy, 38000 Grenoble, France}
\author{Benoit Coasne}
\email{benoit.coasne@univ-grenoble-alpes.fr}
\affiliation{ Univ.\ Grenoble Alpes, CNRS, LIPhy, 38000 Grenoble, France}
\alsoaffiliation{ Institut Laue Langevin, F-38042 Grenoble, France}

\usepackage{pdfpages} 
\usepackage{pgffor} 

\makeatletter
\AtBeginDocument{\let\LS@rot\@undefined}
\makeatother

\def\supplementfilename{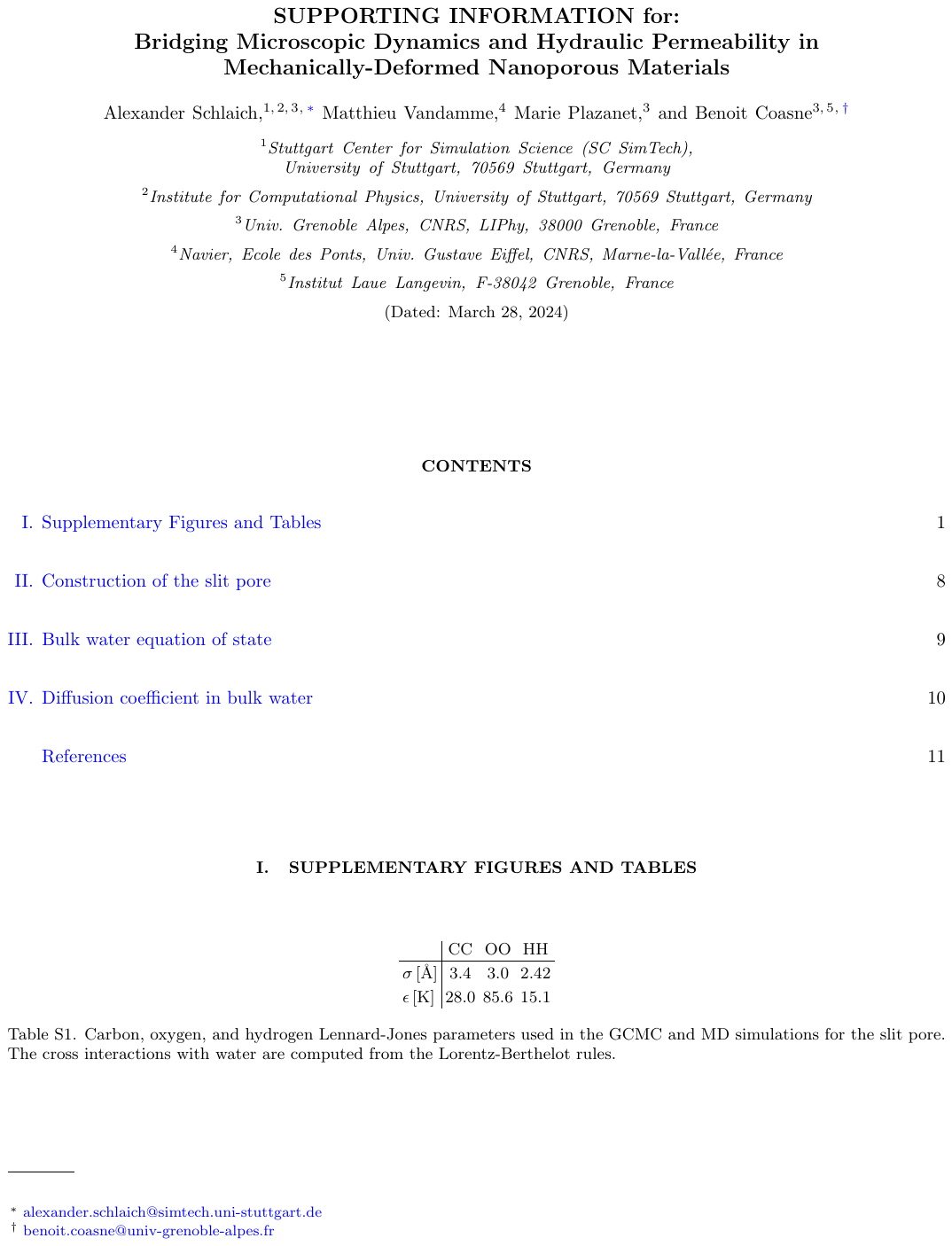}

\pdfximage{\supplementfilename}
\def\numbersupplementpages{\the\pdflastximagepages}

\newif\ifarXiv
\arXivtrue 

\begin{document}

\maketitle

\begin{abstract}
\noindent In the field of nanoconfined fluids, there are striking examples of
deformation/transport coupling in which mechanical solicitation of the confining
host and dynamics of the confined fluid impact each other. 
While this intriguing behavior can be potentially used for practical
applications (e.g.\ energy storage, phase separation, catalysis), the underlying
mechanisms remain to be understood as they challenge existing frameworks.
Here, using molecular simulations analyzed through concepts inherent to
interfacial fluids, we investigate fluid flow in compliant nanoporous materials
subjected to external mechanical stresses. 
We show that the pore mechanical properties significantly affect fluid flow as
they lead to significant pore deformations and different density layering at the
interface 
accounted for by invoking interfacial viscous effects.
Despite such poromechanical effects, we show that the thermodynamic properties
(i.e.\ adsorption) can be linked consistently to Darcy's law for the
permeability by invoking a pore size definition based on the concept of
Gibbs' dividing surface.
In particular, regardless of the pore stiffness and applied external stress, all
data can be rationalized by accounting for the fluid viscosity and slippage at
the interface independent of a specific pore size definition.
Using such a formalism, we establish that the intimate relation
--- derived using the linear response theory --- between collective diffusivity
and hydraulic permeability remains valid.
This allows for linking consistently microscopic dynamics experiments and
permeability experiments on fluid flow in compliant nanoporous materials.
\end{abstract}

\noindent
Fluid/solid interfaces, which are inherent to vicinal and confined liquids as
encountered in nanofluidic devices and nanoporous materials, are host to a
wealth of molecular mechanisms such as adsorption and chemical reactions, but
also to electrokinetic aspects (electrical double layer, crowding,
etc.).\cite{schochTransportPhenomenaNanofluidics2008, bocquet_nanofluidics_2010,
kavokine_fluids_2021,
bazantDoubleLayerIonic2011, laineNanotribologyIonicLiquids2020,
polsterElectricaldoubleLayerRevisited}
With the boost in nanosciences and nanotechnologies, these specific surface
phenomena are already implemented in important applications such as energy
storage, catalysis, lubrication or
depollution.\cite{coppensNatureinspiredApproachReactor2012,
yangCarbonNanotubeMembranes2013, shaoNanoporousCarbonElectrochemical2020,
eijkelNanofluidicsWhatIt2005, huber_soft_2015}
Yet, despite significant progress in our understanding of surface forces and
confinement effects on the thermodynamics and dynamics of fluids, the behavior
of nanoconfined systems still challenges existing frameworks even when simple
liquids are considered.
Among important aspects that remain to be understood,
there are now important experimental and numerical reports on the role of
mechanical deformation on transport of gases and liquids in nanoconfined
environments.\cite{ma_water_2015, bocquet_phonon_2015,
marbachTransportDispersionWiggling2018, marcotteMechanicallyActivatedIonic2020,
nohPhononFluidCouplingEnhanced2022}
Formally, the coupling between mechanics of the confining host and dynamics of
the confined fluid can manifest itself at various scales and in different
fashions. 
Locally, i.e.\ at the fluid/solid interface, momentum transfer between
phonons in the solid phase and molecules in the fluid phase are expected to give
rise to complex surface thermodynamic and dynamical
aspects.\cite{nohPhononFluidCouplingEnhanced2022,
bocquet_phonon_2015, marcotteMechanicallyActivatedIonic2020}
At the pore scale and beyond, mechanical deformation of the pore network
drastically impacts the fluid microscopic dynamics and, in turn, its macroscopic
permeability.\cite{weiMicroscaleInvestigationCoupling2019,
berthonneauMesoscaleStructureMechanics2018,
hoseiniEffectMechanicalStress2009, zhangEffectMechanicalLoad2018,
baiCoupledProcessesSubsurface2000}

Originally, despite the complexity of biological channels and biological objects
in general, biology is an important field in which  fascinating
mechano-transport mechanisms were identified.
Detailed investigation on ionic channels and lipid membranes have unraveled
complex phenomena such as the mechanotransduction response of ionic transfer
upon mechanical or pressure solicitation.\cite{cox_biophysical_2019}
First systematic efforts to stimulate such coupling employed 
man-made nanochannels (carbon nanotubes).\cite{ma_water_2015}
Recently, using single digit experiments on carbon nanotubes,
\citeauthor{marcotteMechanicallyActivatedIonic2020} were able to reproduce
mechanically activated ionic
transport.\cite{marcotteMechanicallyActivatedIonic2020} 
Using 2D nanoporous membranes, \citeauthor{nohPhononFluidCouplingEnhanced2022}
used molecular dynamics simulations displaying a striking impact of membrane
mechanics on water desalination.\cite{nohPhononFluidCouplingEnhanced2022}
Depending on the deformation amplitude and frequency, these authors observed a
drastic effect on water permeability accompanied by a small decrease in salt
rejection. 
By analyzing in detail the dynamics of the deformable membrane and vicinal
water, the vibrational matching between the membrane and water molecules was
found to be the key factor governing the resulting flow.
Despite the important works cited above, several key questions remain left
unanswered.
In particular, while available data point to the phonon--fluid coupling as the
origin of flow modifications, the exact role of mechanical deformation ---
including its quantitative impact --- remains to be assessed.
In particular, the validity of the intricate connection between permeability ---
as defined in macroscopic experiments such as in Darcy’s law --- and the
collective microscopic diffusivity remains to be established when mechanical
solicitation is applied.
Extending such formalism to porous materials and, in particular, to nanoporous
solids is an important step to design mechanical control and stimulation of
fluid adsorption and flow in confined geometries. 

Darcy's law for fluid flow through a porous medium has originally been proposed for macroscopic permeability measurements.\cite{darcy_les_1856} 
Yet, it can be rigorously derived from fundamental conservation laws, i.e.\ the
Navier--Stokes equations.\cite{whitaker_flow_1986,osullivan_perspective_2022}
In detail, in the case of a single-phase fluid considered here (assumed to be
confined in the $z$-direction), one considers the molecular flux 
\begin{equation}
    J =\int_{-L_z/2}^{L_z/2} \rho(z) v(z) \,\mathrm{d}z
    \label{eq:flux}
\end{equation}
induced by a pressure gradient $\nabla P$, where $\rho(z)$ is the local fluid
molecular density and $v(z)$ its corresponding velocity.
The integration boundaries denote a
length-scale $L_z$, which corresponds to the domain size.
If the density is taken to be homogeneous, this molecular flux $J = \rho \bar{v}$
corresponds to a flow rate (mean velocity)
$\bar{v} = -k/\eta \nabla P = -K \nabla P$,
where $\eta$ is the fluid viscosity, $k$ the permeability and $K = k/\eta$ the
(hydraulic) permeance.\cite{muskat_flow_1937,osullivan_perspective_2022}
Direct assessment of $K$  is difficult in general  in
molecular simulations since a constant pressure gradient simulation needs to be
set-up,\cite{thompsonDirectMolecularSimulation1998,
aryaCriticalComparisonEquilibrium2001, bhatiaModelingMixtureTransport2008}
 requiring for explicit treatment of reservoirs and therefore also introducing surface
effects.\cite{martinEffectPressureMembrane2001}
On the other hand, Onsager's relation $J=-D_0/\left(k_\mathrm{B}T\right) \times \nabla \mu$ relates the
flux to the chemical potential gradient $\nabla \mu = -f_x$ in the direction $x$
of the flow ($k_\mathrm{B}$ and 
 $T$ are the Boltzmann constant and temperature, respectively).
The latter situation can be directly considered in molecular simulations through
the measurement of the collective diffusivity $D_0$ as a response to a constant
force $f_x$.
A general approach to relate the transport coefficients due to different driving
forces was developed by
\citeauthor{onsagerReciprocalRelationsIrreversible1931}.\cite{onsagerReciprocalRelationsIrreversible1931}
To relate the above Darcy and Onsager laws, an incompressible liquid is usually
assumed for which we can use straightforwardly the Gibbs--Duhem equation:
$\rho\,\mathrm{d}\mu = \mathrm{d}P$
(which therefore relates the pressure and chemical potential gradients as driving forces).
In so doing, one obtains the following classical result for an incompressible
fluid:\cite{bhatia_molecular_2011}
\begin{equation}
	K = D_0/\left(\rho k_\mathrm{B}T\right).
	\label{eq:permeance_classical}
\end{equation}

Previous studies have considered density-dependence of diffusion,
\cite{falk_subcontinuum_2015, obliger_free_2016,
sastre_molecular_2018, bukowski_connecting_2021}
spatial density and viscosity
heterogeneity,\cite{bhatiaModelingMixtureTransport2008,
schlaichHydrationFrictionNanoconfinement2017} and the the influence of pore
size\cite{chakraborty_confined_2017, tinti_structure_2021} or flexibility on
transport.\cite{nohPhononFluidCouplingEnhanced2022,
marbachTransportDispersionWiggling2018}
Here, we report on a molecular simulation and theoretical approach to
investigate the coupling between mechanical load and flow permeability. Using a simple yet representative model of nanoporous materials, we design a
set-up in which the flow of water induced by a driving force (pressure gradient)
is monitored as a function of the mechanical stress applied in a direction
perpendicular to the pore surface.
To assess the validity of our results, both equilibrium and non-equilibrium
molecular dynamics are employed; using a fluctuation-dissipation approach based
on the Green--Kubo formalism allows verifying that the data and analysis are not
biased due to unphysical coupling between the flow and deformation when
driving forces are applied in non-equilibrium molecular dynamics.
For different pore sizes and mechanical loadings, we analyze the role of the external
mechanical stimulus (stress) and pore mechanical property (stiffness) on the
system’s response --- both the resulting flow and mechanical deformation are
analyzed simultaneously to unravel their interplay.
The advective transport of confined water under mechanical solicitation is then
analyzed through the prism of the fundamentals of interfacial
fluids as developed in the field of nanofluidics and fluids confined in
nanoporous materials.
This allows us to probe the origin of the impact of mechanical deformation on
flow through its effect on the interfacial viscous layer.
Using the linear response theory and the underlying fluctuation-dissipation
theorem, we show that --- akin to non-compliant porous materials --- the
macroscopic permeability can be linked to the microscopic collective diffusivity
as probed using simple molecular simulation.
Beyond verifying its validity for compliant systems, this fundamental relation
 provides a means to define the pore size
that consistently describes both the thermodynamics and dynamics of the confined
fluid. 
In particular, using excess quantities as defined in the Gibbs dividing surface
formalism, our data for the collective diffusivity and permeability are found to
quantitatively match.

\begin{figure*}[htbp]
	\centering
	\includegraphics[width=\textwidth]{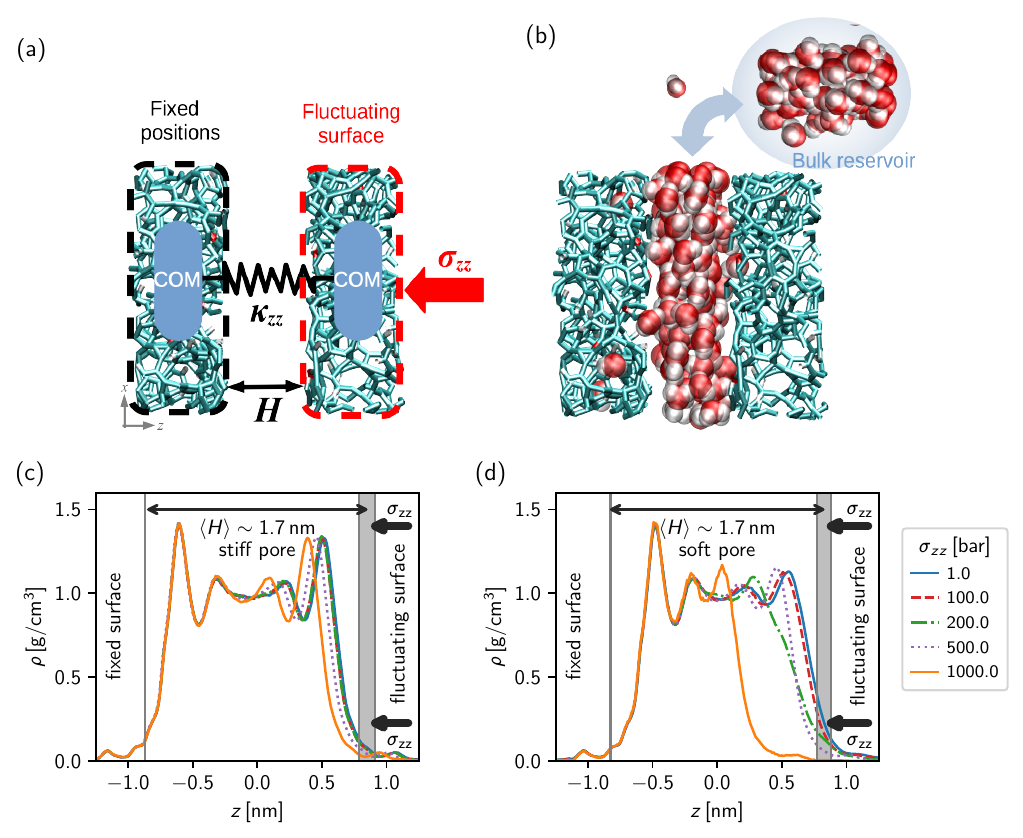}
	\caption
	{{\bfseries Adsorption in compliant nanoporous materials.}
		(a) Set-up of the compliant slit-pore considered in this study.
		The positions of the left pore wall atoms are kept fixed whereas the
		right pore wall is allowed to move as a rigid body.
		The two pores are connected via a Hookean spring of stiffness $\kappa_{zz}$
		acting between their respective centers of mass (COM) and
		several $\kappa_{zz}$ are considered to tune the mechanical pore properties.
		An external stress $\sigma_{zz}$ is applied as a body force acting on  the fluctuating surface.
        For a given spring constant $k$, depending on the chemical potential of
        the external reservoir, the pore gets filled as indicated in (b).
        The resulting surface-to-surface separation fluctuates with an average
        position $\langle H\rangle$ defined as the distance from the planes
        passing through the atom centers at the wall external layer as indicated
        in (a).
        (c) and (d) show water density profiles for a slit pore with
		$\langle H \rangle \sim 1.7 \,\mathrm{nm}$ for a stiff (c) and soft (d) compliant
		material for different applied external stresses $\sigma_{zz}$.
		The chemical potential $\mu$ is fixed to a value corresponding to a
		pressure of about 215 bar to ensure all pores are filled.
		The vertical solid bars, which indicate the positions of the lower and
		upper walls, define the pore spacing $H$ (the left wall is fixed so that
		its position does not fluctuate).
		For $\sigma_{zz}=1000\,\mathrm{bar}$ in (d) one water layer gets
		expelled, leading to a reduction of the pore size from
		$1.6\,\mathrm{nm}$ ($\sigma_{zz} = 500\,\mathrm{bar}$) to
		$1.25\,\mathrm{nm}$.
		}
	\label{fig:fig1}
\end{figure*}

Our setup consists of a compliant slit pore composed out of a realistic dense
and hydrophobic carbon material.\cite{bousigeRealisticMolecularModel2016}
The two pore walls separated by a distance $H$ between the outermost carbon
atoms are treated as a rigid body and connected by a Hookean spring of stiffness
$\kappa_{zz}$ acting between the center of mass (COM) of the pore walls, see
\cref{fig:fig1}(a) and Section II of the Supplementary Information.
While the left wall in \cref{fig:fig1}(a) is fixed in space, the right wall is
allowed to fluctuate freely and to respond to an external stress in
$\sigma_{zz}=f_z N_\mathrm{s}/(L_x L_y)$ in the $z$-direction normal to the
surface (such an external constraint is applied via a force $f_z$ that acts on
all $N_\mathrm{s}$ surface atoms with $L_x = L_y=2.5\,\mathrm{nm}$ the lateral
dimensions of the periodic simulation box).
The harmonic potential $V(l) = \kappa_{zz}(l-l_0)^2$, which depends on the spring extension
$l$, can be related to the slit pore's young modulus,
$E_{zz} = \kappa_{zz} l_0 / (L_x L_y)$, where $l_0$ is the equilibrium length between the
wall's COM resulting in the pore width $H$, see Figure S1 in the Supporting Information.
To mimic water exchange with the bulk external environment as indicated in \cref{fig:fig1}(b), we perform
Grand-Canonical Monte-Carlo (GCMC) and Molecular Dynamics (MD) simulations using the LAMMPS
simulation package, all details are given in \nameref{sec:methods}. 
In line with intrusion experiments in such hydrophobic materials, the water
chemical potential is chosen such that its bulk pressure corresponds to
$P_0 = 215 \,\mathrm{bar}$, see Section III of the Supplementary Information.

We study two systems: (i) a non-compliant pore ($E_{zz}=\infty$) and (ii) a
compliant system, which can swell upon water adsorption.
For the compliant case, we consider two values for the pore modulus, 
$E_{zz}=0.2$ and 2 GPa.
Then, in a second step, for the compliant systems we explicitly address the
impact of pore size fluctuations on water transport through these pores.
In detail, we let the pore relax to its equilibrium size in the presence of
water, but then fix it to its average value $\langle H \rangle$.
In so doing, we keep the thermodynamic state of the confined water and can study the
influence of fluctuating and non-fluctuating pore walls
(by fixing the pore size $H$ to $\langle H \rangle$ or by letting $H$ fluctuate around $\langle H \rangle$).

\subsection{Adsorption and swelling in compliant nanoporous materials}

The density profiles shown in \cref{fig:fig1}(c) and (d) reveal significant
layering at the surface fixed in space (left walls).\cite{israelachviliMolecularLayeringWater1983,
hayashiGrandCanonicalMonte2002, liStructuredViscousWater2007}
Similarly, layering is also observed --- albeit less pronounced --- for the stiff pore
($E_{zz} = 2 \,\mathrm{GPa}$) shown on the right side of \cref{fig:fig1}(c),
where the data shows that the application of an external stress leads to
expulsion of confined water and stronger layering effects.
For the softer pore ($E_{zz} = 0.2\,\mathrm{GPa}$), the data shown on the right
side of \cref{fig:fig1}(d) show that surface fluctuations smear out the layering
at the fluctuating surface.
Upon increasing the stress $\sigma_{zz}$ in the range from $1$ to
$1000\,\mathrm{bar}$, the water density profiles change significantly since the
average pore separation is able to adjust to the water structure.
At $\sigma_{zz} = 1000 \,\mathrm{bar}$, in the case of the soft pore, we observe
a transition in the profiles corresponding to the expulsion of one water layer,
corresponding to a jump in pore size from about $1.6\,\mathrm{nm}$ at
$500\,\mathrm{bar}$ to roughly $1.25\,\mathrm{nm}$ at $1000\,\mathrm{bar}$.
The data shown for that case correspond to an equilibrium length of the spring
of $l_0 = 2.84\,\mathrm{nm}$, which is the largest value considered in this work.

\begin{figure*}[htbp]
	\centering
	\includegraphics[width=\textwidth]{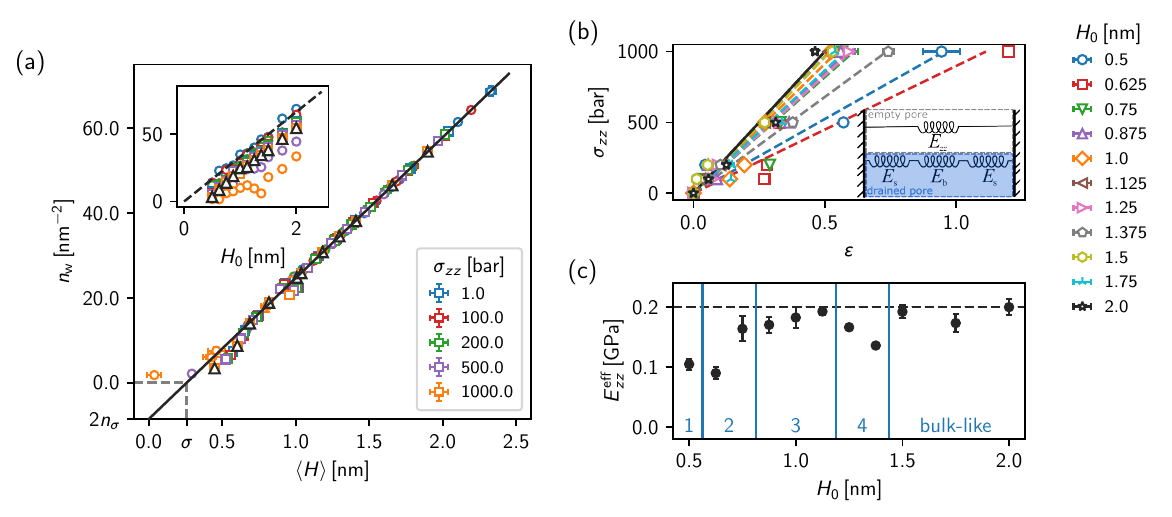}
	\caption
	{{\bfseries Water confinement and pore swelling in compliant nanoporous materials.}
		 (a) Water surface density in molecules/$\mathrm{nm}^2$ as obtained using GCMC
		 simulations for the soft ($E_{zz} = 0.2 \,\mathrm{GPa}$, circles) and
		 stiff ($E_{zz} = 2 \,\mathrm{GPa}$, squares) compliant nanoporous material
		 at different external stresses $\sigma_{zz}$. 
		 Black triangles denote the simulation data for a non-compliant, i.e.\
		 infinitely stiff, nanopore.
		 The black line corresponds to the expected variation based on the bulk
		 water density taking into account excess surface density  $n_\sigma$.
		 The inset shows the same data versus the equilibrium pore size $H_0$  as
		 obtained in the absence of any fluid. The dashed line shows the slope
		 expected from the bulk water density.
        (b) Stress-strain relation for soft pores, $E_{zz}=0.2 \,\mathrm{GPa}$,
        with different nominal pore size $H_0$ as indicated by the legend on
        the right. Dashed lines denote fits of the effective modulus according
        to $\sigma_{zz} = \varepsilon E_{zz}^\mathrm{eff}$.
		The solid line denotes the empty pore
        mechanical response without any effect due to adsorbed water, $\sigma_{zz} =
        \varepsilon E_{zz}$.
		The inset shows the equivalent circuit model as discussed in the text.
        (c) Effective modulus of the soft system for different equilibrium pore
        sizes obtained from the fits in (b).
		The vertical lines denote the different regimes, where in the density
		profiles 1-4 water layers can be observed. Above $\sim
		1.4\,\mathrm{nm}$ the water in the center of the slab is bulk-like,
		\textit{cf.}\ Supplementary Figures 3 and 4.}
	\label{fig:fig2}
\end{figure*}

To further quantify water adsorption in the compliant material, we show in
\cref{fig:fig2}(a) the mean number of water molecules in the slit nanopore normalized to the
lateral area of the pore, $n_\mathrm{w}=N_\mathrm{w}/\left(L_x L_y\right)$.
The dashed line corresponds to the bulk water molecular volume,
$n_\mathrm{w}/\langle H\rangle = 33.2\,\mathrm{nm^{-3}}$.\cite{schlaichHydrationFrictionNanoconfinement2017}
The inset of \cref{fig:fig2}(a) shows the areal water number as a function of
the equilibrium pore size in the absence of any fluid, which follows from the
initial choice of the length of the harmonic spring, $H_0 = l_0 - 0.84
\,\mathrm{nm}$ that connects the walls' COM.
For a non-compliant, i.e.\ frozen, pore (black triangles), the slope corresponding to the bulk density is well reproduced, but with a shift $z_\mathrm{G}$ in the
surface position corresponding to the Gibbs Dividing
Surface.\cite{chattorajAdsorptionGibbsSurface1984} 
In detail, the latter is defined for a single interface via the water surface
excess with respect to its bulk phase,
\begin{equation}
    z_\mathrm{G} = z_a + \int_{z_a}^{z_b}
        \frac{\rho(z_b)-\rho(z)} 
            {\rho(z_b)-\rho(z_a)}\,\mathrm{d}z,
\end{equation}
where $\rho(z_a) \sim 0$ and $\rho(z_b)=\rho_\mathrm{b}$ are the corresponding fluid
densities far from both sides of the interface, i.e.\ no water inside the
solid and water bulk density in the fluid phase.
For the compliant pores (squares and circles), either swelling or compression is
observed depending on the mechanical load applied, cf.\ inset of
\cref{fig:fig2}(a).
All simulation data collapse to a master curve in \cref{fig:fig2}(a) when the 
average effective pore separation $\langle H \rangle$
--- defined as the distance between the outermost surface atoms --- is employed.
The excellent agreement in the slope observed for different surfaces stresses
when large separations are considered reveals that nanoconfined water is nearly
incompressible with a density that sufficiently away from the interface
corresponds to the bulk density.
According to Gibbs adsorption theory, we determine the surface excess
defined for a single interface as
\begin{equation}
    n_\sigma = \int_{-\infty}^{z_\mathrm{G}}
                        \left[\rho(z)-\rho(z_a)\right] \,\mathrm{d}z
                  + \int_{z_\mathrm{G}}^\infty
                        \left[\rho (z)-\rho (z_b) \right] \,\mathrm{d}z.
    \label{eq:Gibbs_excess}
\end{equation}
This can be conveniently done for confined system by fitting fitting the offset
of the line corresponding to bulkd density to all simulation data for which
$\langle H \rangle >1\,\mathrm{nm}$ in \cref{fig:fig2}(a) and yields
$2 n_\sigma = -8.6\,\mathrm{nm^{-2}}$.
Such a negative value indicates that  water is depleted from the hydrophobic
surface. This result was expected for the hydrophobic surface chemistry studied here which leads to non-favorable fluid/solid interactions.
The associated surface separations $\langle H \rangle <\sigma$ at
which the slit pore accommodates no water molecules,
$\sigma=2.65\,\mathrm{\AA}$, perfectly coincides with the
water kinetic diameter (which is related to the mean free path of molecules in a
fluid phase).\cite{ismail_gas_2015}


The change of pore size upon external mechanical stimulation can  be further
analyzed by determining the stress-strain relation in \cref{fig:fig2}(b)
For all systems considered, we find the strain
$\varepsilon = (H_0-\langle H\rangle )/H_0$
to be roughly related linearly to the normal stress $\sigma_{zz}$,
where here, per convention, a shrinkage corresponds to a positive strain.
This is not necessarily the expected behavior since the interfacial  structure
and compressibility of water are non-negligible for the small pores considered.
However, the strain-stress linearity  
allows us to extract the apparent (effective) modulus of the slit pore under
drained conditions, $E_{zz}^\mathrm{eff} = \sigma_{zz}/\varepsilon$,
in analogy to the apparent tangent drained bulk modulus derived from a
generalized poromechanics framework.\cite{brochard_poromechanics_2012}
The corresponding equivalent circuit depicted in the inset of \cref{fig:fig2}(b)
consists of the mechanical properties of the empty pore, characterized through
the modulus $E_{zz}$ acting in parallel with the water mechanical response under
drained conditions.
In fact, the latter can be further dissected into the water bulk and surface
contributions, $E_\mathrm{b}$ and $E_\mathrm{s}$, respectively, acting in
series (as the total strain upon application of a normal stress $\sigma_{zz}$
must be equal to the sum of the strains of the individual springs). 
The corresponding effective modulus follows as
\begin{equation}
	{E_{zz}^\mathrm{eff}} = {E_{zz}} + \frac{1}{
		E_\mathrm{b}^{-1} + 2E_\mathrm{s}^{-1}}.
	\label{eq:Ezz}
\end{equation}

If we assume a linear relation to hold, linear fits to the simulation data in
\cref{fig:fig2}(b) allow for the determination of the effective Young modulus
in presence of water, 
$E_{zz}^\mathrm{eff} = \sigma_{zz}/\varepsilon$.\cite{gor_adsorption-induced_2017,
kolesnikov_models_2021}
Such data are shown in \cref{fig:fig2}(c) for the soft compliant pore with
$E_{zz}=0.2\,\mathrm{GPa}$.
For sufficiently large separations $H_0\gtrsim 1.5\,\mathrm{nm}$, interfacial effects
(at sufficiently large strain, see \cref{fig:fig2}(b)) become negligible if
the water is allowed to exchange with the reservoir. In that case,
$E_{zz}^\mathrm{eff} \sim E_{zz}$ is recovered, i.e.\ the response to an applied
stress of the filled pore system corresponds to that of the confining material.
Since the water bulk modulus is about
$25\,\mathrm{kbar}$,\cite{motakabbir_isothermal_1990}
this implies that the term $2E_\mathrm{s}^{-1}$ dominates the denominator in
\cref{eq:Ezz}, i.e.\ the surface layer modulus must be small such that the total
fraction vanishes.
Further analysis of the data shown in \cref{fig:fig2}(c) for the soft pore and
in Figure S2 in the Supplementary Information for the stiff pore reveals that
$E_{zz}^\mathrm{eff} \lesssim E_{zz}$.
Since $E_\mathrm{b}$ is positive and independent of the pore width, this
indicates that $E_\mathrm{s}$ depends on $H_0$ and $E_{zz}$, and is negative.
Although this is a counterintuitive result at first (since it indicates a
negative apparent surface compressibility), it is in line with generalized
poromechanics stating that the apparent tangent bulk modulus can be smaller than
its counterpart in bulk depending on the adsorption
properties.\cite{brochard_poromechanics_2012}
Physically, it corresponds to the concept of a disjoining
pressure,\cite{eskandari-ghadi_mechanics_2021} which depending on the wetting
properties either is attractive or repulsive.\cite{kanduc_water-mediated_2016}
Note, that the typical signature of hydration forces --- the oscillatory stress
vs.\ pore size curve --- is encapsulated in the effective modulus since the pore
can mechanically adapt to the molecular water structure.

Interestingly, deviations for the smaller pores with $E_{zz}^\mathrm{eff} \neq E_{zz}$
can be correlated with the density layering shown in
\cref{fig:fig1} (Figures S3 and S4 in the Supplementary Information
show the same data for all systems).
The vertical blue lines in \cref{fig:fig2}(c) indicate the corresponding number
of water layers.
For $H_0 \gtrsim 1.5\,\mathrm{nm}$, the water density profile in the slit pore center
becomes homogeneous and equal to its bulk value (so that density heterogeneities
correspond to two distinct layers at the solid/fluid interface).
Depending on the pore separation $H$, in the case of the smaller pores, water
organizes into one, two, three or four layers. Moreover, significant deviations from
the prescribed modulus can be observed. 
The exact influence of the interfacial contribution depends sensitively on the
equilibrium surface separation $\langle H \rangle$ of the system in contact with the reservoir.
It also depends on the applied mechanical stress $\sigma_{zz}$ which strongly affects the
thermodynamic state of nanoconfined water. Indeed, even if confined water in each system is at
the same chemical potential $\mu$ and temperature $T$, its thermodynamic state
also depends on the prescribed mechanical condition $\sigma_{zz}$.

\subsection{Transport in compliant nanoporous materials}

\noindent \textbf{Collective diffusivity.}
Having assessed the thermodynamic and adsorption behavior of water
in compliant slit nanopores, we now turn to  transport in the
presence of a fluctuating interface.
We first focus on Onsager flow, where the universal thermodynamic driving force is
a gradient in the chemical potential, $\nabla \mu$.
Solving \cref{eq:flux} for the mean velocity $\bar{v}$ while taking into account
the heterogeneous density distribution $\rho(z)$ yields
\begin{equation}
    \bar{v} = \frac{\int \rho(z) v(z)
    \,\mathrm{d}z^\prime}{\int \rho(z) \,\mathrm{d}z^\prime} =
    -\frac{D_0}{k_\mathrm{B}T} \nabla \mu 
    \label{eq:vbar}
\end{equation}
where  Onsager's relation defined in 
\cref{eq:permeance_classical} was used.
\cref{eq:vbar} rigorously defines $D_0$ from the velocity $v(z)$ and density 
$\rho(z)$ profiles without any need of defining a pore size $H$ or homogeneous density
$\rho$.
This is a significant advantage in the definition of $D_0$, which can be probed
experimentally --- e.g.\ using quasi-elastic coherent neutron scattering.
In contrast, the permeance $K$ defined in Darcy's equation relies on the
ambiguous definition of a pore size as discussed below.

\begin{figure}[htb]
	\centering
	\includegraphics[width=\textwidth]{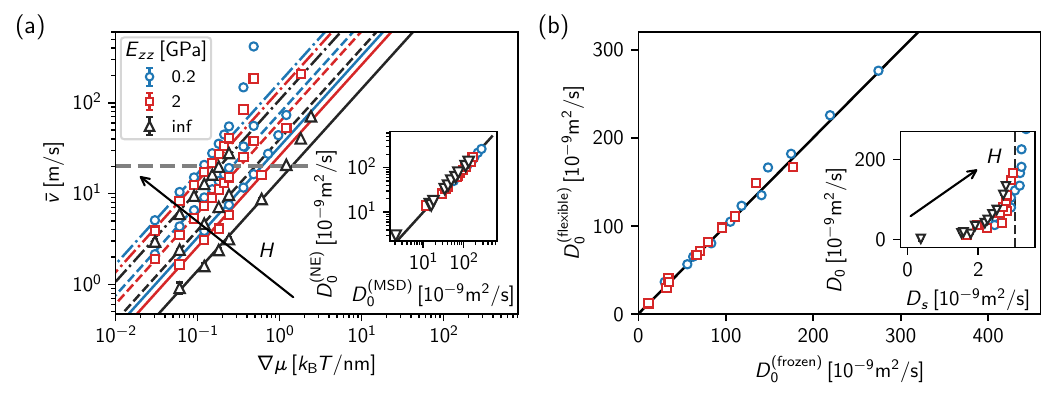}
	\caption
	{{\bfseries Probing dynamics in equilibrium and non-equilibrium conditions.}
		 (a) Typical fits for the mean water velocity $\langle v\rangle$ vs.\ driving force $\nabla \mu$.
		 Color code corresponds to the stiffness as shown in legend.
		 Data are shown for three characteristic pore sizes $H \approx 0.75 \,\mathrm{nm}$ (solid lines),
		 1.25 nm (dashed lines) and 1.75 nm (dashed-dotted lines).
		 The inset shows the collective diffusion coefficient determined from the linear response
		 regime of the mean water velocity (non-equilibrium molecular dynamics) vs.\ the collective 
		 diffusion coefficient determined from the mean-squared displacement (equilibrium molecular dynamics).
		 All values for the flexible pores are shown for $\sigma_{zz} = 1 \,\mathrm{bar}$.
		 (b) Diffusion coefficient in flexible pores vs.\ frozen pores (same mean pore size and water number but non-fluctuating walls).
		 The black line indicates the bisector for clarity.
		 The inset shows the collective diffusion constant $D_0$ vs.\ the self-diffusion coefficient $D_\mathrm{s}$ to display the role of collective contributions (cross-terms) in diffusion.
		 The vertical dashed line corresponds to the bulk value
		 $D_\mathrm{s}^{\mathrm{(b)}}=3.03\,\mathrm{m^2/s}$ determined from
		 independent simulations with finite size scaling, see Section IV of the
		 Supplementary Information for details.
	}
	\label{fig:fig3}
\end{figure}

In the following, we unravel the effect of pore elasticity by
limiting the discussion to the specific case where the external stress is set to
$\sigma_{zz} = 1\,\mathrm{bar}$.
\Cref{eq:vbar} provides a means to directly determine $D_0$ from non-equilibrium
simulations, where a chemical potential gradient in the $x$-direction parallel
to the pore is applied via a homogeneous
force field acting on each water molecule, $\nabla \mu = -f_x$, see
\nameref{sec:methods} for details.
\Cref{fig:fig3} shows the resulting mean velocity which can be obtained directly
from the density and velocity profiles according to \cref{eq:vbar}. In the
case of point particles, the flow rate  can be further simplified to
$\bar{v} = 1/\tau \int_0^\tau \mathrm{d}t \, 1/N_\mathrm{w}
\sum_{i=0}^{N_\mathrm{w}} v_i(t)$, where $v_i(t)$ denotes the instantaneous
velocity of a water molecule's center of mass in the direction of the flow at
time $t$ and the integral averages over the simulation time $\tau$. 

\Cref{fig:fig3}(a) shows $\bar{v}$ as a function of the driving force $f_x = -\nabla \mu$ for
different pore widths $\langle H \rangle$ and stiffnesses $E_{zz}$.
Whereas \cref{eq:vbar} is intrinsically a linear relation, the linear response regime  is not 
achieved in the molecular simulations (as well as in experiments) when too large driving forces are used.
However, we find  that for sufficiently small driving forces the
linear response limit can be probed [lines in \cref{fig:fig3}(a)].
In particular, the interfacial friction between water and the surface governs
the onset of nonlinear effects. As a result,  the  critical driving force at the
crossover between the linear and nonlinear regimes sensitively depends on the
surface separation as well as the pore
flexibility.\cite{schlaichHydrationFrictionNanoconfinement2017}
The frictional stress at the interface
increases with $\bar{v}^3$  within the classical Darcy-Weisbach theory.
This explains the emergence of a critical value
$\bar{v}_\mathrm{crit} = 20\,\mathrm{m/s}$ which is rather independent of the pore size
and stiffness. 
Such a velocity cutoff for the linear response regime can be rationalized
with an activated diffusion of water molecules over barriers at the
interface.\cite{erbasViscousFrictionHydrogenBonded2012}
Note that, however, the corresponding driving forces are highly dependent on the
pore size, which is in line with the expectation from classical Hagen-Poiseuille
theory.

In practice, we fit $D_0^\mathrm{(NE)}$ in \cref{eq:vbar} to the non-equilibrium
simulation data in \cref{fig:fig3}(a) for velocities 
$\bar{v} < \bar{v}^\mathrm{crit}$.
The fits are indicated as black and color lines, whereas the horizontal dashed
gray line in \cref{fig:fig3}(a) denotes the employed value
$\bar{v}^\mathrm{crit} = 20\,\mathrm{m/s}$.
As shown in the inset of \cref{fig:fig3}(a), the obtained values are in
perfect agreement with the equilibrium molecular dynamics results relying on the
mean-squared displacement and the fluctuation-dissipation theorem, see
\nameref{sec:methods} and Figures S7 and S8 in the Supplementary Supplementary Information.
As can be observed from the data in \cref{fig:fig3}(a), the collective diffusion $D_0$
increases with pore size in line with a Poiseuille flow, but also with pore
flexibility (black via red to blue data).
To shed further light on the dependence $D_0(E_{zz})$, we performed an additional
set of simulations at constant (non-fluctuating) pore size with a value fixed at
$H = \langle H \rangle$, where the brackets denote the time-average of the
fluctuating pore size for the specific stiffness $E_{zz}$.
The resulting diffusion coefficients in \cref{fig:fig3}(b) perfectly agree
between the fluctuating and non-fluctuating pores with the same fixed water
number $N_\mathrm{w}$.
This result points to the fact that $D_0$ depends on adsorption/layering rather
than on pore size fluctuations. 
However, we note that the rather large pore size employed here together with the
incompressibility of water and the employment of a stiff surface can explain the
negligible impact of pore size fluctuations.
In other words, for smaller pores (or, equivalently, larger surface to volume
ratios), pore size fluctuations and, in general, mechanical deformations are
expected to affect liquid flow in nanoconfinement.
This point will be addressed further below.

The collective diffusivity as defined in Onsager's law, i.e.\
\cref{eq:flux,eq:vbar}, differs strongly from the self-diffusion coefficient
$D_\mathrm{s}$ that follows from tracking a tagged
molecule.\cite{hansenTheorySimpleLiquids2013}
The collective correlations in a fluid are usually positive and thus strongly
enhance $D_0$ compared to $D_\mathrm{s}$ (and in the dilute limit $D_0 \to
D_\mathrm{s}$).
The inset of \cref{fig:fig3}(b) shows the corresponding relation $D_0$ vs. 
$D_\mathrm{s}$ for a confined system, for details on the calculation see \nameref{sec:methods}.
Interestingly, due to interfacial depletion/adsorption effects, $D_\mathrm{s}$
in a confined system can also become larger/smaller than its counterpart in bulk
at the same chemical potential/pressure,
$D_\mathrm{s}^{(0)}=3.03 \,\mathrm{m^2/s}$, which we determined in independent
simulations for our water model, see Section IV of the Supplementary Information.
A detailed discussion of such effects is out of scope of this work and well
represented in text books.\cite{kargerDiffusionNanoporousMaterials2012}

\begin{figure*}[htbp]
	\centering
	\includegraphics[width=1.0\textwidth]{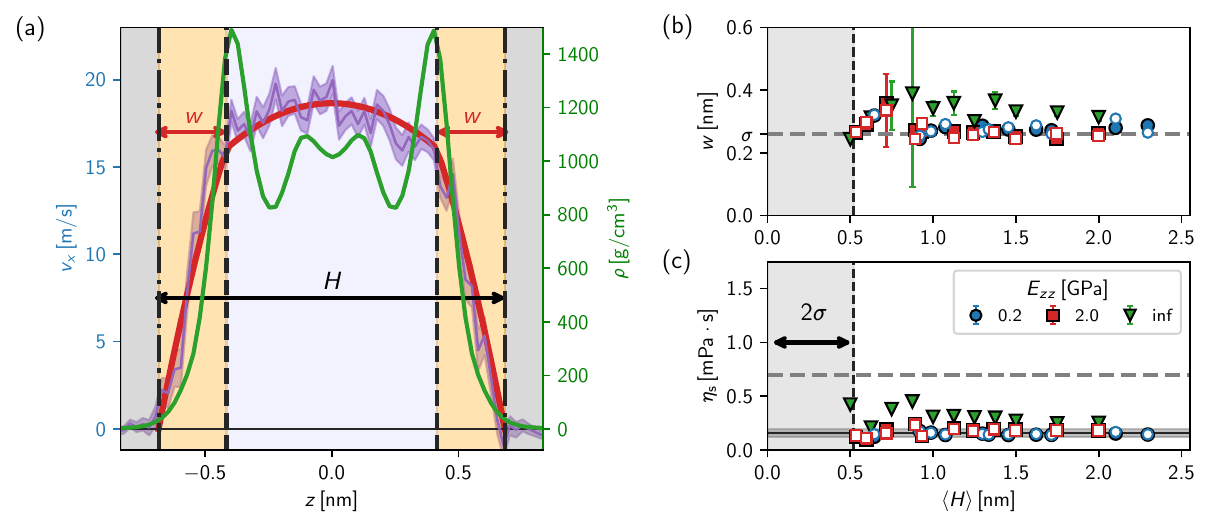}
	\caption
	{{\bfseries Velocity profiles and interfacial properties.}
		(a) Velocity (purple line) and density (green line) profiles in the
			pore-center coordinate frame (see text) obtained in the linear
			regime for the slit pore with $\langle H \rangle= 1.4 \,\mathrm{nm}$
		    (indicated by the dash-dotted vertical black lines) and pore
		    stiffness $E_{zz} = 2 \,\mathrm{GPa}$.
            The red line denotes the fit against a two-zone Poiseuille flow
            accounting for an interfacial viscous layer, c.f.\
			\cref{eq:twozone}.
            The yellow shaded area denotes the fitted interfacial width $w$,
			the shaded area of the velocity profile the standard error of the
			mean.
        (b) Fitted values of the interfacial width and (c) viscosity for the
	        different pore stiffnesses and sizes.
		The dashed horizontal line in (b) denotes the value
		$\sigma=0.26\,\mathrm{nm}$ obtained via the surface excess in
		\cref{fig:fig2}(a).
		The dashed horizontal line in (c) denotes the bulk value of the water
		model used in the simulations, $\eta_\mathrm{b} = 0.695\,\mathrm{mPa\cdot s}$,
		whereas the solid line and shaded area denote the averaged viscosity in
		the interfacial layer,
		$\eta_\mathrm{s} = 0.16 \pm 0.04 \,\mathrm{mPa \cdot s}$.
		The filled symbols in (b) and (c) correspond to fluctuating pores
		whereas the empty symbols are obtained for frozen pores,
		which nearly perfectly overlap with the data for the fluctuating pores.
		For $\langle H \rangle\leq 2\sigma$, the viscous behavior is purely interfacial as
		indicated by the shaded areas in (b) and (c).
	}
	\label{fig:fig4}
\end{figure*}

\noindent \textbf{Velocity profiles.}
Further insight can be obtained from analyzing the velocity and density
profiles.
To this end, we perform a coordinate transformation from the lab frame
[\cref{fig:fig1}(a)] into the pore frame, i.e.\ $z \to z-z_\mathrm{COM} (t)$,
where $z_\mathrm{COM} = \sum_i [z_i(t) m_i] / \sum_i m_i$ is the fluctuating
center of mass of the pore and the sum runs over all $i$ pore wall atoms.
The resulting density profiles, which are illustrated for the stiffer system and
an average pore size $\langle H \rangle=1.4\,\mathrm{nm}$ in
\cref{fig:fig4}(a), reveal that in this frame
the density profiles are symmetric as expected.
Furthermore, as already noted above, the first peak of the density profile is
strongly depleted from the position of the interface (defined by the outermost
position of the wall atoms).
The velocity profile [purple data in \cref{fig:fig4}(a)] is parabolic in the
slab center and decays nearly linearly to zero at the pore wall, $z=H/2$.
Importantly, the velocity profiles are perfectly in line
with the solution of the Stokes flow in the presence of an interfacial fluid
film of thickness $w$.
The modified Poiseuille profile accounting for continuity of the velocity and
stress at the interface between the two zones is given by
\begin{align}
	v_x(z) &= \frac{\nabla P}{2} \left[\frac{1}{\eta_\mathrm{b}}
		\left( a^2 - z^2\right) 
		+ \frac{1}{\eta_\mathrm{s}} \left(\frac{H^2}{4} - a^2\right) 
	\right]  &\mathrm{for}& \quad 0 \leq |z| \leq \frac{a}{2} \nonumber \\
	v_x(z) &= \frac{\nabla P}{2 \eta_\mathrm{s}}
				\left[  \frac{H^2}{4} - z^2\right]
				&\mathrm{for}& \quad \frac{a}{2} \leq |z| \leq H,
	\label{eq:twozone}
\end{align}
where we introduced the width of the bulk-like region, $2 a = H - 2 w$ to
simplify notation and $\eta_\mathrm{b}$ and $\eta_\mathrm{s}$ denote the bulk
and surface layer effective viscosities, respectively.
To apply \cref{eq:twozone} to our simulations involving a
gradient in the chemical potential instead of the pressure, we make use of
the Gibbs--Duhem relation for incompressible water ($\textrm{d}P = \rho \textrm{d}\mu$) to obtain
$\nabla P = - f_x N_\mathrm{w} / \left(L_x L_y H\right)$.

Fits of \cref{eq:twozone} to the simulation data shown in \cref{fig:fig4}  
reveal excellent agreement with the two-zone Poiseuille flow (see also
Figures S3 and S4 in the Supplementary Information for all systems).
These data show a rather universal value for the interfacial width
$w \approx \sigma = 0.26 \,\mathrm{nm}$, c.f.\ \cref{fig:fig4}(b).
To improve numerical stability during the fitting procedure, we fix $w=\sigma$
in the following discussion. 
This approach is justified since it leads to comparable
root-mean-square error in the fits to the velocity profiles.
The interfacial layer has a strictly reduced viscosity as shown in \cref{fig:fig4}(c).
As can be seen in \cref{fig:fig4}(a), water is strongly depleted from the surface
within the width $w$. This is believed to  result in a reduced value of $\eta_\mathrm{s}$ that is
quasi independent of pore stiffness and size for the compliant pores with a
value $\bar{\eta_\mathrm{s}} = 0.16\pm 0.04 \,\mathrm{mPa\,s}$ [solid black line
and corresponding shaded area in \cref{fig:fig4}(c)].
Importantly, the results of the fitting procedure are independent of whether we
consider fluctuating or non-fluctuating pores at the same pore size and water content
[empty symbols in \cref{fig:fig4}(b,c)].
Noteworthy, in the non-compliant nanopores, $E_{zz} = \infty$, i.e.\ when the pore
cannot adapt to the water structure by any modification of the pore size, water
layering is enhanced (Figure S3 in the Supplementary Information) and the interfacial viscosity
converges only slowly for comparatively large pores $>2\,\mathrm{nm}$ towards
$\bar{\eta_\mathrm{s}}$.

As observed from \cref{fig:fig4}(a), both the density and velocity profiles
approach zero at the steric definition of the pore size $H$ and the usual
\textit{no-slip} boundary conditions perfectly holds.
In the pore center, at a distance larger than $\sigma=0.26\,\mathrm{nm}$ from the
surface, the velocity profile is parabolic by construction from
\cref{eq:twozone}.
It is noteworthy that this parabolic region of width $H-2\sigma$ is different
from the thermodynamic pore size $H-\sigma$ as defined above from the Gibbs
dividing plane.
Since both the density and velocity decay quickly to zero at distances smaller
than $\sigma$ from the surface, this region barely contributes to the flux
according to \cref{eq:flux}.
Thus, it is tempting to introduce a coarse-grained model of a simple planar
Poiseuille flow in the region $[-(H/2-\sigma),(H/2-\sigma)]$.
Two questions naturally arise at this stage. First, what is the corresponding boundary
condition at the pore surface, $v\left(\pm(H/2-\sigma)\right)$?
Second, what is the contribution of the deviations from the parabolic profile in
the vicinity of the surface to the overall flow?
Since the fluid velocity does not vanish at the interface to the bulk-like flow
region, the \textit{no-slip} boundary condition does not hold. 
In this case, a \textit{slip} boundary condition is introduced instead with a
characteristic slip length 
\begin{equation}
	b = \pm v_x(z_\mathrm{surf}) / \left( \mathrm{d}v_x / \mathrm{d}z \right)_{z_\mathrm{surf}}.
	\label{eq:slip}
\end{equation}
Defining $b$ requires the knowledge of the position of the corresponding surface
$z_\mathrm{surf}$, where the hydrodynamic boundary condition is applied. 
There is no absolute natural  choice for $z_\mathrm{surf}$  in the case of an
atomistically resolved surface.\cite{botanHydrodynamicsClayNanopores2011}
This has led  to different choices in the literature ranging from definitions based
on the shear-stress correlation function,\cite{bocquet_hydrodynamic_1994,
herrero_shear_2019} the use of the Gibbs' dividing surface, the repulsive contribution of
the surface interaction,\cite{sokhan_fluid_2002} or simply based on the
position of the surface atoms.\cite{geng_slip_2019}
All these definitions sensitively influence the numerical value inferred for the
slip length $b$.

Based on the excellent description obtained above with the two-zone Poiseuille
flow accounting for interfacial viscosity for the simulated velocity profiles,
we here propose the effective \textit{hydrodynamic} diameter $H-2\sigma$.
In so doing, accounting for slippage and an effective viscosity reproduces the
bulk-like region of the simulation results. 
Such a coarse-graining approach --- despite the question about its validity to be
addressed below --- significantly simplifies the analytical treatment of the flow
profiles.
In detail, the Poiseuille flow with slip is given by 
\begin{equation}
	v_x(z) = \nabla P \frac{{H^\star}^2} {8 \eta} \left[ 
		1 + \frac{4 b}{H^\star} - \left(\frac{2 z}{H^\star}\right)^2
	\right],
	\label{eq:Poiseuille_slip}
\end{equation}
where the asterisk denotes the --- generally unknown --- effective pore size.
We now identify $H^\star=H-2\sigma$ and consider only the velocities in the
region $[-(H/2-\sigma),(H/2-\sigma)]$ --- where the
profiles shown in Fig. \ref{fig:fig4} and for all systems in Figures S3 and S4
in the Supplementary Information are parabolic.
Concerning the width $w$ of the interfacial layer,
\cref{eq:slip,eq:Poiseuille_slip} define the slip length
$b = \eta_\mathrm{b} / \left(2 \eta_\mathrm{i}\right) \times 
	\left(Hw - w^2\right) / \left(H/2 - w\right)$.
With this definition, using the values $w = \sigma$ and $\bar{\eta_\mathrm{s}}$
for large $H$ results in $b = 1.2 \pm 0.3 \,\mathrm{nm}$, cf.\ Figure S5 in the
Supplementary Information.

Importantly, to obtain the above expression for $b$, we have redefined the pore
size as the effective hydrodynamic pore size according to $H^\star = H-2\sigma$.
It is worth mentioning, that if the pore size $H^\star = H$ would be employed
and only the parabolic part of the velocity profile were considered, both
the slip length and the apparent viscosity
$\eta_\mathrm{app} = \eta_\mathrm{b} + 2w/H \times 
	\left( \eta_\mathrm{i} - \eta_\mathrm{b} \right)$ 
become dependent on $H$.\cite{vo_transport_2016}
Depending on the interfacial properties, $\eta_\mathrm{app}$ can be orders of
magnitude larger than the bulk
viscosity,\cite{ma_origins_2015, liStructuredViscousWater2007,
schlaichHydrationFrictionNanoconfinement2017,
leng_fluidity_2005} or significantly decrease,\cite{shaat_wettability_2020,
kohler_size_2016} as observed also for the hydrophobic surfaces considered here.
In summary, our observations from  the density and velocity profiles reveal the
two following points.
(1) The thermodynamic pore size as defined from Gibbs's dividing surface differs
	from a steric definition $H$ by a the characteristic size of a water molecule,
	$\sigma = 2.65 \,\mathrm{\AA}$.
(2) At the length of a molecular size, one observes interfacial viscous effects,
	i.e.\ the pore size where bulk hydrodynamic behavior is found is $H-2\sigma$.
	Note that this is only partially equivalent to claiming that the
	hydrodynamic pore size is $H-2\sigma$ because one needs to take into account
	slippage.
In such a coarse-grained model --- besides the fact that in general the pore size
cannot be rigorously defined such that the apparent viscosity needs to be fitted
--- one misses a  part in the flux due to neglecting the interface layer.
This contribution to the total flux is in general rather small as will be
discussed below.

\begin{figure}[htbp]
	\centering
	\includegraphics[width=\textwidth]{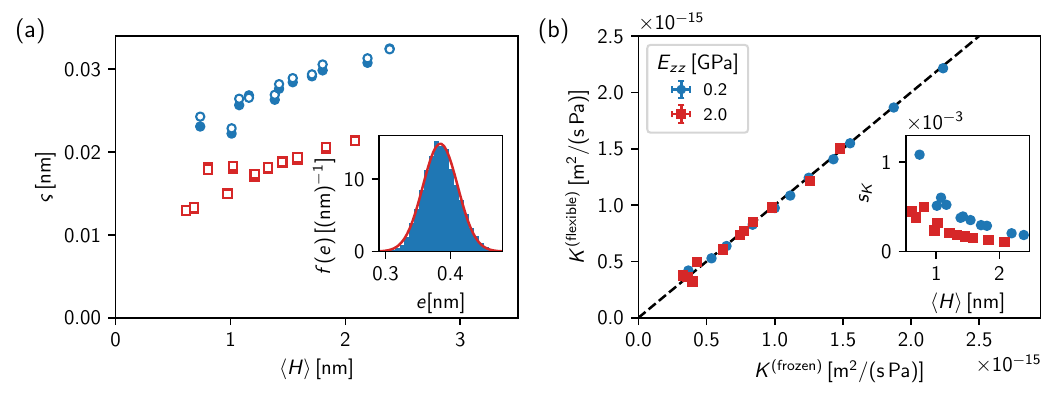}
	\caption
	{{\bfseries Influence of fluctuating vs.\ non-fluctuating pore surfaces.}
		 (a) Variance of the pore size fluctuations in equilibrium (empty
		 symbols) and non-equilibrium simulations (full symbols).
		 The inset shows an example  of the pore extension distribution function
		 $f(e)$ for $E_{zz} = 0.2 \,\mathrm{GPa}$ and $H_0 = 1\,\mathrm{nm}$,
		 red line denotes a fit of a Gaussian to the data.
		 (b) Permeance of the compliant, fluctuating pores vs.\ the respective
		 value in a rigid pore at the same average pore size and particle
		 number.
		 The inset shows the expected relative permeability enhancement
		 $s_K$ due to pore size fluctuations (see text).
	}
	\label{fig:fig5}
\end{figure}

\noindent \textbf{Pore size fluctuations.} 
To address the underlying effects of pore size fluctuations in a flexible nanopore,
we assume in the following that the pore size of a pore with spring equilibrium
length $H_0$ filled with water can be described using a Gaussian distribution
around the mean spring extension $e$ given by
$f(e; E_{zz})={1}/{\left(\varsigma \sqrt{2\pi}\right)}
	\exp{\left[-{1}/{2}\left(\left(e-\bar{e}\right)/\varsigma\right)^2\right]}$
(the extension $e$ is due to swelling and the variance is $\varsigma$).
Note that we dropped the dependence on the external stress since in the
following  we limit our discussion to $\sigma_{zz} = 1\,\mathrm{bar}$.
For all systems, we find that the fluctuations are well described by this
mathematical form for $f(e)$, cf.\ inset of \cref{fig:fig5}(a).
Dropping the dependence on the modulus, the average pore size follows as 
$\langle H \rangle = H_0 + \int_{-\infty}^\infty e f(e)\,\mathrm{d}e$.

To ensure a compact presentation of the following discussion, we here employ the
coarse-grained model presented above. 
In other words, the water velocity profiles in compliant
pores can --- neglecting interfacial effects ---
be modelled by a Hagen-Poiseuille law when accounting for slip and
effective viscosity, see \cref{eq:Poiseuille_slip}.
An extension to the case where the interfacial viscous effects are treated
explicitly is straightforward.
A further simplification that is  often made in the analysis of Darcy
flows is to treat the water density $\rho = N_\mathrm{w} / V_\mathrm{w} =
N_\mathrm{w}/(A\langle H\rangle)$ as homogeneous ($V_\mathrm{w}=A\langle
H\rangle$ is the volume accessible to the water and $A=L_x L_y$ is the lateral
area of the slit pore).
Noting that  the density profiles presented in \cref{fig:fig4} and for all
systems in Figures S3 and S4 in the Supplementary Information
are not homogeneous, we will discuss the limitations and impact of this simplification below.
The permeance of a fluctuating nanopore with a Poiseuille velocity profile and
slip follows from \cref{eq:Poiseuille_slip} as
\begin{equation}
	K = \frac{\langle H^\star \rangle^2}{4\eta}\left(\frac{2b}{\langle H^\star \rangle} + \frac{1}{3}\right).
	\label{eq:permeance_poiseuille_slip}
\end{equation}
As discussed above, our NEMD simulations reveal that
the slip length quickly approaches its asymptotic constant value
$b=1.2\,\mathrm{nm}$ for sufficiently large nanopores.
Moreover, using a consistent definition of the pore size, the viscosity $\eta =
\eta_\mathrm{b}$ is found to be independent of $\langle H\rangle$ and pore
fluctuations.
Furthermore, employing $f(e)$ to express the mean pore size, we combine
\cref{eq:vbar,eq:permeance_poiseuille_slip} and perform the
Gaussian integral, such that the collective diffusion coefficient can be
expressed as
\begin{equation}
	D_0 = k_\mathrm{B}T \langle\rho K\rangle 
		= \frac{k_\mathrm{B}T}{6\eta}\frac{N_\mathrm{w}}
			{A \left( 1+\frac{2\sigma}{\langle H^\star \rangle} \right)}
		 \langle H^\star \rangle
			\left( 1 + \frac{3b}{\langle H^\star \rangle} \right),
	\label{eq:D_0-gaussian}
\end{equation}
which yields the important result of $D_0$ being independent of pore size
fluctuations, in good agreement with the results presented in
\cref{fig:fig3}(b).
Although being a first order approximation neglecting any fluid density
heterogeneities or changes in the effective viscosity and slip,
\cref{eq:D_0-gaussian}
captures remarkably well the qualitative behavior observed in the present work 
(see gray dashed line in Figure S6 in the Supplementary Information).

Plugging the pore size distribution using $f(e)$ into \cref{eq:flux} allows us 
further to define the average flux according to
$J=-\langle \rho v \rangle = - N_\mathrm{w} /
			\left( A \langle v / H \rangle \right)$.
This directly yields the permeance for a fluctuating slit pore,
\begin{equation}
	K = -\frac{A}{N\nabla \mu} \langle v H \rangle.
	\label{eq:permeance_twist}
\end{equation}
For a non-fluctuating pore 
$\langle v H \rangle = \langle v \rangle \langle H \rangle$
which yields the classical result 
$K^\mathrm{(frozen)} = D_0/\left(k_\mathrm{B}T\right) \times A\langle H\rangle / N$. 
The latter expression relates the permeance and collective diffusion via the thermal
energy and fluid density $\rho = N/\left(A\langle H \rangle \right)$.
On the contrary, \cref{eq:permeance_twist} reveals that a coupling between pore size
fluctuations and the fluid velocity can impact transport in compliant pores (in contrast to the collective diffusivity which is expected to be independent of fluctuations as shown above).
For the Poiseuille flow, the coupling can be made quantitative by calculating
the second moment of the Gaussian integral appearing in
\cref{eq:permeance_poiseuille_slip}, resulting to leading order in 
$K \sim H_0^2 + \int_{-\infty}^{\infty} e^2 f(e) \,\mathrm{d}e 
	= \langle H \rangle^2 + \varsigma^2$,
i.e.\ under these assumptions the fluctuations enhance transport in a fluctuating
slit pore. 

We show in \cref{fig:fig4}(a) the variance $\varsigma$ of the pore size
fluctuations determined both from equilibrium simulations, i.e.\ in the absence
of flow, as well as from simulations with applied driving force $\nabla \mu$.
Notably, EQMD and NEMD simulation results agree perfectly, revealing that flow
does not affect the fluctuations in the linear response regime.
Furthermore, fluctuations are in the range of about 
$(1-3) \times 10^{-2} \,\mathrm{nm}$ for the pore sizes 
$H \approx (0.5-2.5)\,\mathrm{nm}$ studied here, i.e.\ fluctuations are about two
orders of magnitude smaller than the average pore size, which we attribute to
the near-incompressibility of water under these thermodynamic and transport conditions.
The resulting permeance $K^\mathrm{(flexible)}$ obtained from NEMD simulations
with fluctuating pores is plotted in \cref{fig:fig4}(b) against the values
$K^\mathrm{(frozen)}$ obtained from simulations with the pore size fixed to the
mean pore size $H=\langle H \rangle$
and same water number $N_\mathrm{w}$ (i.e.\ the positions of the surfaces were fixed).
As expected from the small values of $\varsigma$, the permeance $K$ is quasi
independent of the fluctuations. 
To quantify the expected flow enhancement due to pore fluctuations, we define the coefficient
$s_K = \left( K^\mathrm{(flexible)} - K^\mathrm{(frozen)} \right) / K^\mathrm{(frozen)}
	= \left( \langle H \rangle^2 + \varsigma^2 \right) / \langle H \rangle^2$
shown in the inset of \cref{fig:fig5}(b).
We find values $s_K$ as small as $10^{-3}  - 10^{-4}$ for water in the studied
slit pores, revealing that the impact of pore fluctuations on permeance are
negligible in that case.
However, for stronger fluctuating pores --- and, hence, more compressible
liquids --- the discussed effects are expected to be more pronounced.

\begin{figure}[htbp]
	\centering
	\includegraphics[width=1.\textwidth]{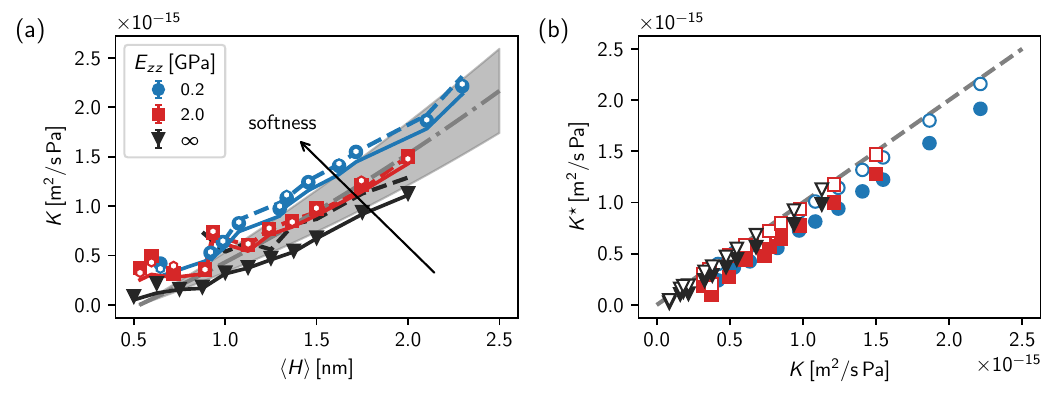}
	\caption
	{{\bfseries Permeance.}
		The permeance $K$ is shown as a function of pore size $\langle H \rangle$ for the soft
		(blue circles), stiff (red squares), and rigid pores (black triangles).
		Filled and empty symbols denote simulation data evaluated according to
		\cref{eq:permeance_twist} for the fluctuating and frozen pores,
		respectively.
		The gray dashed-dotted line denotes the expected permeance from Poiseuille
		equation \cref{eq:permeance_poiseuille_slip} if a homogeneous density profile, 
		bulk viscosity, and a slip length of $1.2\pm0.3\,\mathrm{nm}$ are assumed (see
		Figure S5 in the Supplementary Information). 
		Shaded area denotes the corresponding uncertainty in the extrapolated
		slip length.
		Colored dashed lines further take into account the fitted slip lengths
		for each pore height and stiffness, data $H<1\,\mathrm{nm}$ are excluded
		due to their large uncertainty, see Figure S5 in the Supplementary
		Information.
		Solid lines are constructed from the fitted Poiseuille profiles taking
		into account the density weighted velocity profiles,
		\cref{eq:permeance_fit}.
		(b) Permeance $K^\star$ calculated according to \cref{eq:Kstar} vs.\ the
		permeance measured directly from the flux in (a). Solid symbols
		correspond to the classical result of \cref{eq:permeance_classical},
		neglecting the surface excess $n_\sigma$, whereas empty symbols take
		into account the fitted surface excess $n_\sigma=-8.6\,\mathrm{nm^{-2}}$
		obtained in \cref{fig:fig2}.
		The dashed line serves as guide to the eye.
	}
	\label{fig:fig6}
\end{figure}

\noindent \textbf{Flow enhancement in compliant nanopores.} 
Our simulation data allow us to quantify the effect of the compliance of a
nanoporous material on the resulting permeance according to \cref{eq:permeance_twist}.
In \cref{fig:fig6}(a), we plot the permeance $K$ as a function of the resulting average pore
size $\langle H \rangle$.
Comparison of the fluctuating compliant pores with simulations at the same mean
pore size and water numbers [filled vs.\ empty symbols in \cref{fig:fig6}(a)]
shows that fluctuations in the slit nanopore do not
enhance transport, also see \cref{fig:fig5}(b) and in line with the findings
discussed in the previous paragraph.
Yet, the molecular simulation data show a strong increase of $K$ with decreasing stiffness
at the same average pore size $\langle H \rangle$.
In other words, softer pores lead to transport enhancement due to the pore
compliance itself.
This is due to the fact that at the same bulk pressure (i.e.\ chemical
potential), due to the different elastic properties of the slit pores,
water can accommodate in different density profiles, and thus a larger net
amount of water is present that can be transported, c.f.\
\cref{fig:fig1,fig:fig2}.
These thermodynamic effects that change the water density profile and viscosity
close to the surface seem to be the main driving parameter leading to flow
properties (\cref{fig:fig4}).

The solid lines in \cref{fig:fig6}(a) show the permeance predicted from the two-zone
Poiseuille flow weighted by the water density profile,
\begin{equation}
    K = - \frac{\int\rho(z)v(z)\,\mathrm{d}z}{\int\rho(z)\,\mathrm{d}z} \frac{\langle{H}\rangle A}{N \nabla \mu}.
    \label{eq:permeance_fit}
\end{equation}
In this equation, the term
$\left(\langle H \rangle A\right) / \left(N \nabla \mu\right) = 1/\nabla P$
again relates chemical potential and pressure gradients as driving force
according to a Gibbs--Duhem relation.
The simulation data are excellently described by \cref{eq:permeance_fit} with a
systematic underestimation of the permeance that increases with increasing
softness and pore width.
This is  due to the limited ability of the two-zone Poiseuille flow model to
appropriately describe the profiles, see Figures S3 and S4 in the Supplementary
Information.
While more sophisticated models for the interfacial viscosity --- like e.g.\ an
exponential decay to the interfacial value
\cite{schlaichHydrationFrictionNanoconfinement2017} --- can be established, their
discussion lies out of scope of the present work and the observed deviations do
not change our main conclusions. 

Following the idea of a coarse-grained model for the pore flux above, we now
assume that a Poiseuille flow with slip,
\cref{eq:permeance_poiseuille_slip}  describes accurately the permeance for
$\eta=\eta_\mathrm{b}$ and $b=1.2\pm0.3\,\mathrm{nm}$.
The resulting permeance is shown as dashed gray line and corresponding shaded area in
\cref{fig:fig6}(a), revealing an overall good agreement with the simulation data 
(symbols).
However, the applicability of this model is limited by the fact that the
extrapolated slip length is subject to rather large uncertainty and is not
necessarily the same for the different pore elasticities considered.
Indeed, we find that good agreement using the coarse-grained model with the
measured permeance for $H\gtrsim1\,\mathrm{nm}$ when the distance-dependent
values of $b$ are determined according to \cref{eq:slip} from fits of the
two-zone model as shown in Figure S5 in the Supplementary Information.
For smaller pore sizes, the fitting procedure is not robust enough to accurately
predict the ratio $b/H$.
Significant deviations in the predicted permeance from this approach are only
observed for the pore with infinite stiffness.
This is due to the fact that density layering is the strongest for this system,
which is included in \cref{eq:permeance_fit} and thus shows perfect agreement
with the simulation data [solid lines in \cref{fig:fig6}(a)].
The observations above suggest that for sufficiently large pores the assumption
of a homogeneous density and parabolic Poiseuille profile with slip boundary
conditions predicts accurately the permeance for nanopores (even though
significant interfacial layering and slip effects are present).
Note that this conclusion still relies on introducing a hydrodynamic pore size
$H^\star=H-2\sigma$ as discussed above.

Finally, we now link the permeance --- which is a macroscopic 
quantity amenable to simple experimental testing --- to the microscopic quantity $D_0$.
For a homogeneous fluid, one expects \cref{eq:permeance_classical} to hold.
In \cref{fig:fig6}(b), we compare the permeance $K^\star$
predicted from \cref{eq:permeance_classical} and the collective diffusivity
$D_0$ determined from the simulations (Figure S6 in the Supplementary
Information) with the measured permeance $K$.
These data are compared for the different values $E_{zz}$ considered in this
work.
\cref{eq:permeance_classical} strictly underestimates the permeance due to the
neglect of the surface excess when assuming a homogeneous, bulk-like density.
This can be accounted for by a modified expression,
\begin{equation}
    K^\star = D_0 H \frac{1}{\kbT \left(
        n_\sigma + \rho_\mathrm{b} H\right)},
    \label{eq:Kstar}
\end{equation}
in which we use $n_\sigma = -8.6\,\mathrm{nm^{-2}}$
(see discussion above). The corresponding comparison corresponds to the  empty
symbols in \cref{fig:fig6}(b).
The collapse of the data on the bisector reveals that \cref{eq:Kstar} allows for
a consistent link between $K$ and $D_0$ through the concepts of interfacial
thermodynamics (including the significant surface layering which leads to strong fluctuations with respect to a simple homogeneous density profile).
The latter finding thus paves the way to consistent interpretation of
microscopic dynamics --- incorporated in $D_0$ and measured experimentally
typically in scattering experiments --- to the permeance $K$, that is accessible
from experimental hydraulic flow measurements through a porous material.

In conclusion, whereas the compliance of a nanoporous material strongly
influences the transport in deformable nanopores due to the different density
layering and, thus, different interfacial viscous effects, pore size
fluctuations do not enhance the flow in the system considered here.
While permeance enhancement is expected from a theoretical viewpoint, the
absence of such poromechanical effect is due to the large pores and/or fluid
incompressibility involved in our systems.  
We note, however, that for smaller pores and more flexible materials these
effects could become highly relevant and encanced transport due to the compliant
material appears.
More importantly, we find that a consistent pore size definition can be employed to relate accurately the thermodynamics and transport of the nanoconfined fluid. In particular, when invoking the Gibbs dividing surface, we find that the
classical description through a Poiseuille flow holds perfectly at the
nanoscale. Active driving of fluctuating membranes has been suggested to increase
selectivity and transport,\cite{marbach_active_2017} whereas 
additional effects appear through the phonon-fluid coupling.\cite{nohPhononFluidCouplingEnhanced2022}
Thus, it is promising to further explore flow and selectivity manipulation 
for stronger fluctuating systems in future work. In this context, by providing a consistent description to account for transport and thermodynamics of fluids confined in nanoporous materials, we believe that the present work offers a robust framework for such perspectives.

\subsection{Methods}
\label{sec:methods}

\subsubsection{GCMC/MD Simulations}

All molecular simulation results were obtained using the Large-scale
Atomic/Molecular Massively Parallel Simulator (LAMMPS) version 30 Oct
2019.\cite{plimptonFastParallelAlgorithms1995}
Water is treated using the SPC/E model\cite{berendsenMissingTermEffective1987}
kept rigid via the SHAKE algorithm
\cite{ryckaertNumericalIntegrationCartesian1977} and the slit pore was
constructed as described in Section II of the Supplementary Information.
Input files are freely available in the data repository of the University of
Stuttgart (DaRUS).\cite{darus-3966_2024}
In the molecular simulations, the carbon, oxygen and hydrogen atoms of the host matrix were
treated as Lennard-Jones particles with previously established parameters that
well account for experimental adsorption
isotherms,\cite{bousigeRealisticMolecularModel2016,
billemontAdsorptionCarbonDioxide2013, pikunicStructuralModelingPorous2003}
see Table S1 in the Supporting Information.
All molecular simulations were run at temperature $T=300\,\mathrm{K}$ using a Nosé-Hoover
thermostat\cite{shinodaRapidEstimationElastic2004} with a characteristic damping
time of $0.2\,\mathrm{ps}$. The integration timestep of $2\,\mathrm{fs}$ was
used.
Short-range interactions were cutoff and shifted to zero at $9\,\mathrm{\AA}$ and
long-range electrostatics was treated using the PPPM
method\cite{plimptonParticleMeshEwaldRRESPA1997} with an accuracy of $10^{-4}$.
For the slab system, we additionally used the correction by Yeh and
Berkowitz\cite{yehEwaldSummationSystems1999} with a total vacuum layer of 3
times the slab separation.

During the GCMC/MD simulations, we performed every $1\,\mathrm{ps}$  $2\times10^4$
insertion/deletion steps for water molecules in the slab.
To obtain convergence, typical $10^6$ GCMC steps/water molecule were employed,
which corresponds to a total simulation time of about $10\,\mathrm{ns}$ per
system to sample the equilibrium particle number.
The water chemical potential was converted into a corresponding reservoir
pressure using independent simulations of bulk water, as explained in Section
III of the Supplementary Information.
The value $\mu=-11.4\,\mathrm{kcal/mol}$ used here was
thus found to correspond to an external pressure of $P_0=215\,\mathrm{bar}$.
This value is sufficient for the pores to always be filled within the range of
values $\sigma_{zz}, H_0$ reported in \cref{fig:fig2}.

\subsubsection{Equilibrium MD Simulations}

To setup equilibrium MD simulations, slit pores with the average
number $N_\mathrm{w}$ as determined above were constructed. All other
simulation parameters were left unchanged.
Systems were equilibrated for 1 ns and then productions runs of 250 ns length
were performed during which the unfolded coordinates were recorded every 1 ps to
calculate the mean-squared displacement. The latter were used to infer the collective and self
diffusivities as 
$D_0 = \lim_{t \to \infty} 1/(2 N_\mathrm{w} d t)
	\langle \sum_{i,j} \left[ \mathbf{r}_i(t) - \mathbf{r}_i (0)\right]
	             \cdot \left[ \mathbf{r}_j(t) - \mathbf{r}_j (0)\right] \rangle$
and
$D_\mathrm{s} = \lim_{t\to\infty} 1/(2dt) \langle \left| 
	\mathbf{r}(t) - \mathbf{r}(0) \right|^2 \rangle$,
respectively. In these equations, $d=2$ is the dimensionality of the slit pore
and the corresponding displacement $\mathbf{r}$ is calculated in the planar
directions only.
Typical curves for the mean-squared displacement are shown in Figure S7 in the
Supplementary Information.
Care has to be taken in the determination of the diffusion coefficients from the
mean-squared displacement since on the one hand the limit $t\to\infty$
has to be approximated, but on the other hand the statistics gets very poor for
large $t$ (see Figure S7 in the Supplementary Information).
In practice, to reliably determine a reasonable fitting range, we determine the
time-dependent values $D_0(t) =  1/(2 N_\mathrm{w} d t)
\langle \sum_{i,j} \left[ \mathbf{r}_i(t) - \mathbf{r}_i (0)\right]
			 \cdot \left[ \mathbf{r}_j(t) - \mathbf{r}_j (0)\right] \rangle$
and
$D_\mathrm{s}(t) = 1/(2dt) \langle \left| 
\mathbf{r}(t) - \mathbf{r}(0) \right|^2 \rangle$,
shown in Figure S7 in the Supplementary Information, and average the data over the range where it is
roughly constant, i.e.\ typically between 250-5000~ps to evaluate $D_\mathrm{s}$
and 100-1000~ps to evaluate $D_0$ for all systems
except the smallest pores, where the plateau is observed even for smaller time
intervals only due to the small values of the diffusion coefficients.

Alternatively, the diffusion constant can be determined from the velocity
autocorrelation functions,
$D_0 = 1/(N_\mathrm{w} d) \int_0^\infty \langle \sum_{i,j}
	\mathbf{v}_i(t) \cdot \mathbf{v}_j(0) \rangle$
and 
$D_\mathrm{s} = 1/d \int_0^\infty 
	\langle\mathbf{v}(t) \cdot \mathbf{v}(0) \rangle$.
Note that the velocity correlation function of water is dominated by the
femtosecond timescale as shown in Figure S8 in the Supplementary Information, thus requiring a high dump
frequency for the velocities.
In practice, we recorded the velocities every timestep and after 1 ns the
correlation function was evaluated via fast Fourier transformation using the
convolution theorem.
The resulting correlation functions are averaged over the total simulation time
and finally integration is performed.
The limit $t\to\infty$ needs to be replaced by some appropriate fitting range to
obtain the extrapolation as shown in Figure S8 in the Supplementary Information.
As also shown therein, both approaches give -- as expected -- results that are in perfect
agreement.
However,  $D_\mathrm{s}$ involves an average over all $N_\mathrm{w}$
water molecules whereas $D_0$ corresponds to the water's center of mass
movement. Therefore,  significantly less statistics is used to measure $D_0$.

\subsubsection{Non-equilibrium MD Simulations}

Non-equilibrium simulations were performed by applying a constant force
$f_x=-\Delta \mu / L_x$ on all water oxygen atoms along the $x$-direction.
This driving force corresponds to a chemical potential difference $\Delta \mu$
over the corresponding length $L_x$ of the simulation box.
We made sure that the resulting mean velocity and permeance remained within the
linear response regime, cf.\ \cref{fig:fig3}.
To avoid any coupling between the applied force and the system thermalization,
only the $y$ and $z$ components of the velocity were considered for calculating
the thermal energy.
All other simulation parameters were the same as explained above.
After an initial equilibration of 1 ns, molecular simulations were performed for a total
length of 500 ns per system during which positions and velocities were recorded
every 10 ps.

\subsubsection{Data analysis}

All obtained simulation data were analyzed using
MDAnalysis\cite{michaud-agrawalMDAnalysisToolkitAnalysis2011}
and our in-house open source framework MAICoS,
\url{https://www.maicos-analysis.org/}.
Analysis scripts are freely available on the data repository of the University
of Stuttgart DaRUS, \url{https://doi.org/10.18419/darus-3966}.\cite{darus-3966_2024}

\section{Associated content}

\subsection{Supporting Information}

The Supporting Information is available free of charge on the publisher's website.
\par\noindent
\begingroup
\leftskip4em
	Table S1 and Figures S1-S8,
	details on the construction of the slit pore,
	SPC/E bulk water equation of state for the employed simulation setup,
	diffusion coefficient in bulk water and
	analysis of pore size and particle number fluctuations.
	\par
\endgroup
	
\section{Author Information}

\subsection{Author Contributions}

AS and BC designed the research project; AS performed the simulations; AS
and BC analyzed the simulation data; all authors were involved in preparing
the manuscript.

\subsection{Notes}

The authors declare no competing financial interest.

\section{Acknowledgments}

We acknowledge funding from the Agence Nationale de la Recherche (Project TWIST No.
ANR-17-CE08- 0003) and computation time on the GRICAD infrastructure supported
by the Equip@Meso project (No. ANR-10-EQPX-29-01).
AS acknowledges funding from the DFG under Germany's Excellence Strategy - EXC
2075 – 390740016 and SFB 1313 (Project Number 327154368) and support by the
Stuttgart Center for Simulation Science (SimTech).

\bibliography{twist}

\ifarXiv


\section*{Supplementary material}

# Bridging Microscopic Dynamics and Hydraulic Permeability in Mechanically-Deformed Nanoporous Materials

Alexander Schlaich,[1, 2, 3, *] Matthieu Vandamme,[4] Marie Plazanet,[3] and Benoit Coasne[3, 5, †]

[1]*Stuttgart Center for Simulation Science (SC SimTech),*
*University of Stuttgart, 70569 Stuttgart, Germany*

[2]*Institute for Computational Physics, University of Stuttgart, 70569 Stuttgart, Germany*

[3]*Univ. Grenoble Alpes, CNRS, LIPhy, 38000 Grenoble, France*

[4]*Navier, Ecole des Ponts, Univ. Gustave Eiffel, CNRS, Marne-la-Vallée, France*

[5]*Institut Laue Langevin, F-38042 Grenoble, France*

(Dated: March 28, 2024)

## CONTENTS

## I. SUPPLEMENTARY FIGURES AND TABLES

|  | CC | OO | HH |
|---|---|---|---|
| $\sigma$ [Å] | 3.4 | 3.0 | 2.42 |
| $\epsilon$ [K] | 28.0 | 85.6 | 15.1 |

Table S1. Carbon, oxygen, and hydrogen Lennard-Jones parameters used in the GCMC and MD simulations for the slit pore. The cross interactions with water are computed from the Lorentz-Berthelot rules.

* alexander.schlaich@simtech.uni-stuttgart.de
† benoit.coasne@univ-grenoble-alpes.fr

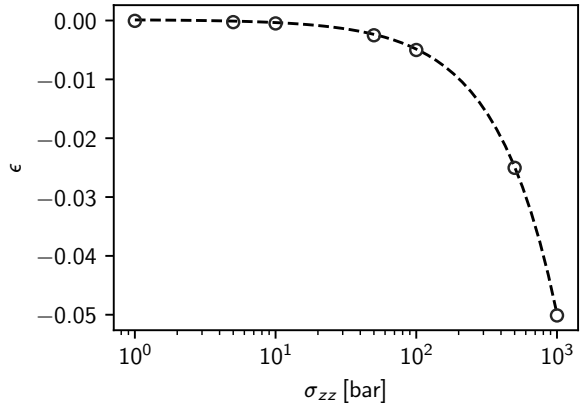

FIG. S1. **Hookean behavior of the empty compliant pore.** Normal stress $\sigma_{zz}$ vs. strain $\epsilon = (l-l_0)/(l_0)$. Symbols denote simulation data for an empty pore, $H = 1\,\mathrm{nm}$ and modulus $E_{zz} = 2\,\mathrm{GPa}$. Dashed line shows the linear behavior, $\epsilon = -\sigma/E$.

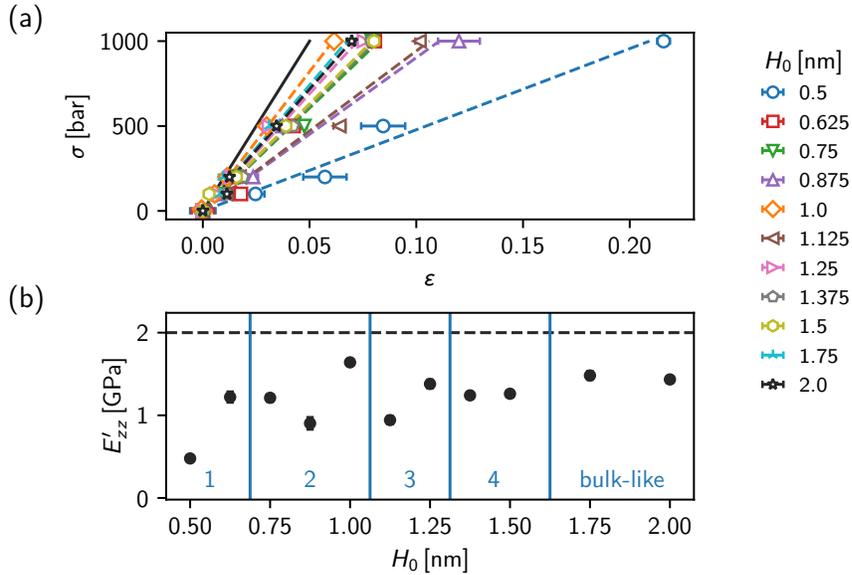

FIG. S2. **Analysis of the effective pore modulus for the stiffer pore,** $E_{zz} = 2\,\mathrm{GPa}$. (a) Stress-strain relation for stiff pores, $E_{zz} = 2\,\mathrm{GPa}$, with different equilibrium pore size $H_0$ as indicated by the legend on the right. Dashed lines denote fits of the effective modulus according to $\sigma_{zz} = \varepsilon E_{zz}^{(\mathrm{eff})}$. The solid line denotes the empty pore mechanical response without any effect due adsorbed water, $\sigma_{zz} = \varepsilon E_{zz}$. (b) Effective modulus of the stiff system for different equilibrium pore sizes obtained from the fits in (a). The vertical lines denote the different regimes, where in the density profiles 1-4 water layers can be observed. Above $\sim 1.7\,\mathrm{nm}$ the water in the center of the slab is bulk-like, *cf.* density profiles shown in Figs. S3 and S4.

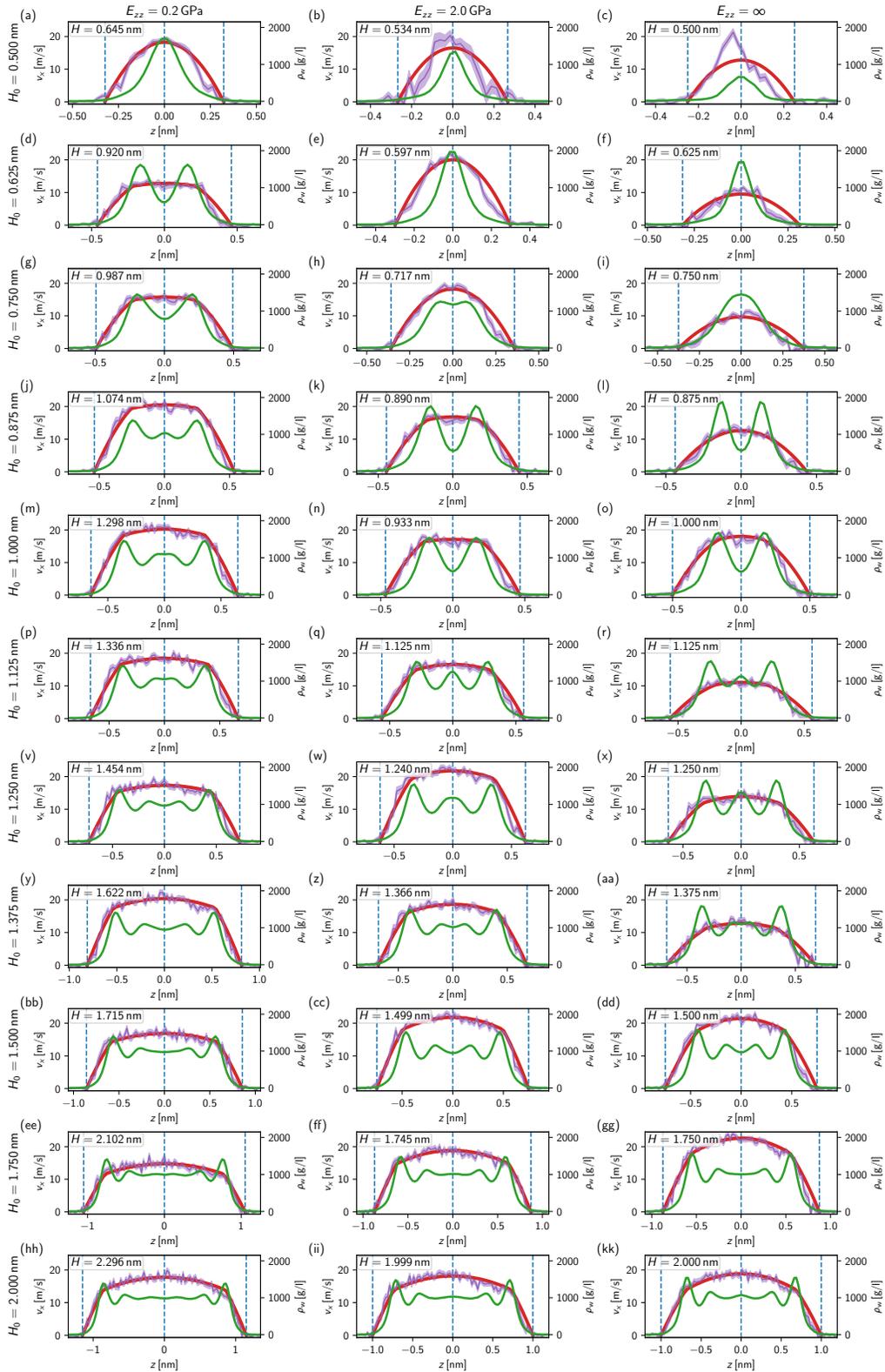

FIG. S3. **Density and velocity profiles for all systems considered.** For the compliant and non-compliant pores we show the density (green) and velocity profiles (purple). Red lines denote fits of the two-zone Poiseuille model to the velocity profiles, see main text. Dashed horizontal lines denote the position of the pore surface and its center, respectively.

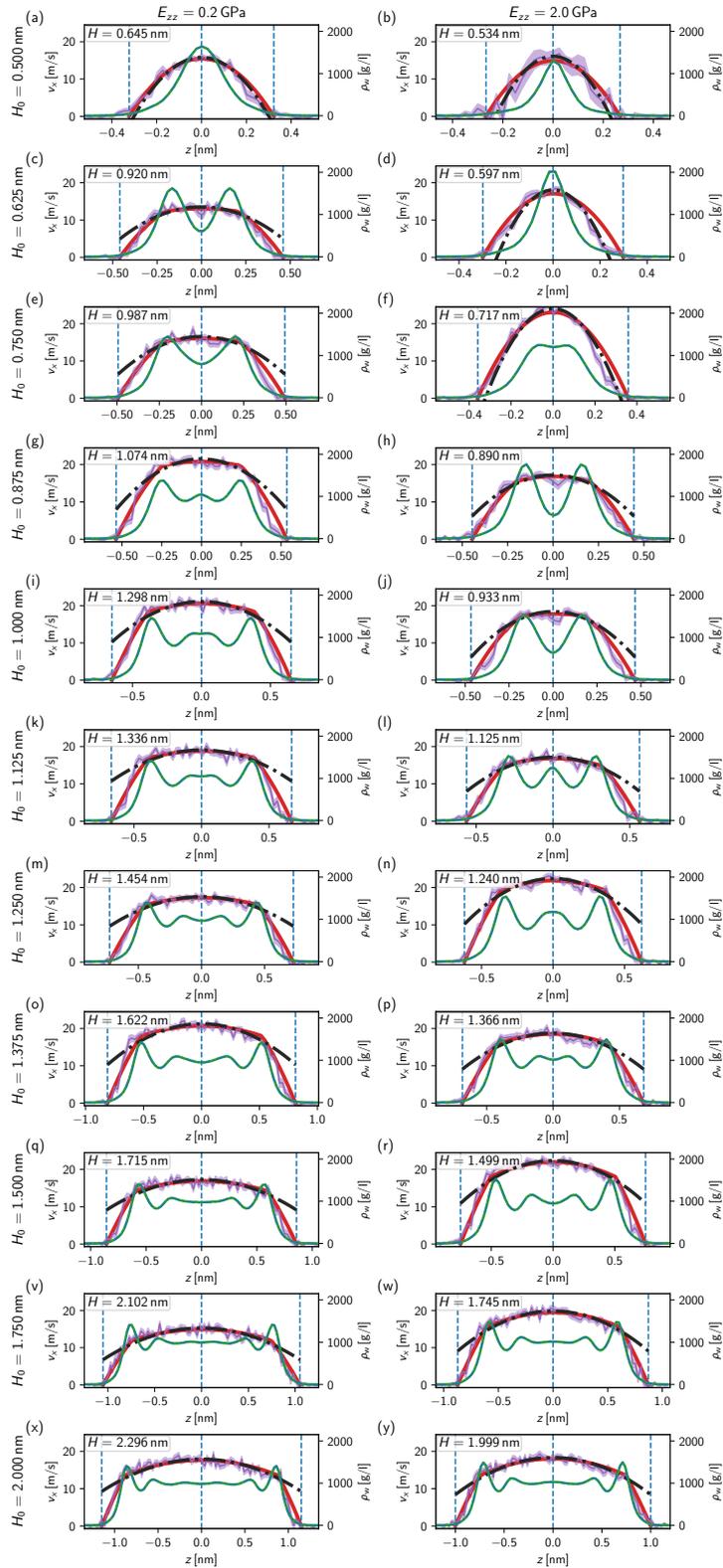

FIG. S4. **Density and velocity profiles the non-fluctuating compliant pores.** As in Fig. S3, the density (green) and velocity profiles (purple) are shown. Red lines denote fits of the two-zone Poiseuille model to the velocity profiles, see main text. Additionally, dahed-dotted black lines indicate the coarse-grained fit of the Poiseuille profile data revealing slip. Dashed horizontal lines denote the position of the pore surface and its center, respectively.

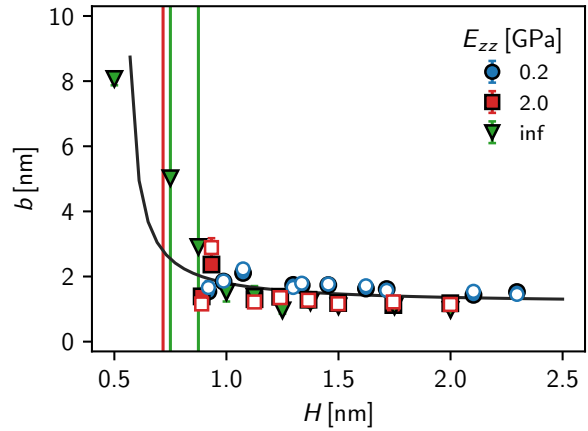

FIG. S5. **Slip length.** Slip lengths according to Eqs. (8) and (9) obtained from the fits of the two-zone Poiseuille flow to the simulation data. The solid line denotes the slip length computed using the values at large distances, $w = \sigma$ and $\bar{\eta}_{\parallel}$. The colored vertical lines denote the error bars, revealing that for the small pores this procedure does not allow to reliably extract the slip length. However, these values are well consistent with the expected behavior from the large pore sizes (solid black line).

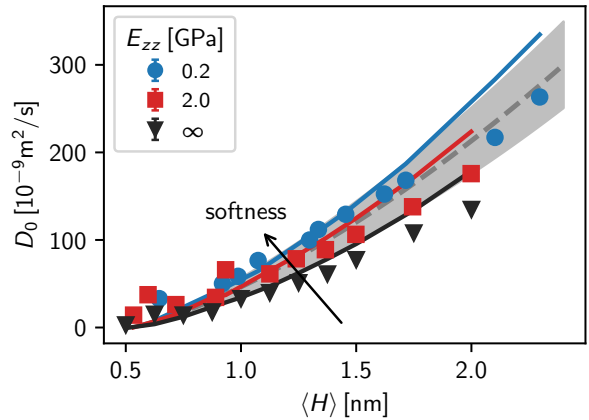

FIG. S6. **Collective diffusion.** The diffusion constant $D_0$ is shown as a function of pore size $H$ for the soft (circles), stiff (squares), and rigid pores (triangles). Symbols denote simulation results from the non-equilibrium simulations according to the data shown in Fig. 3, whereas the dashed gray lines is showing the classical Poiseuille prediction, Eq. (11) of the main text with $\eta = \eta_b$ and $b = 1.2 \pm 0.3$ nm (the shaded are corresponds to the uncertainty in the slip length). Colored lines denote data using Eq. (11) of the main text, but with distance-dependent values of the slip length and apparent viscosity (Fig. S5).

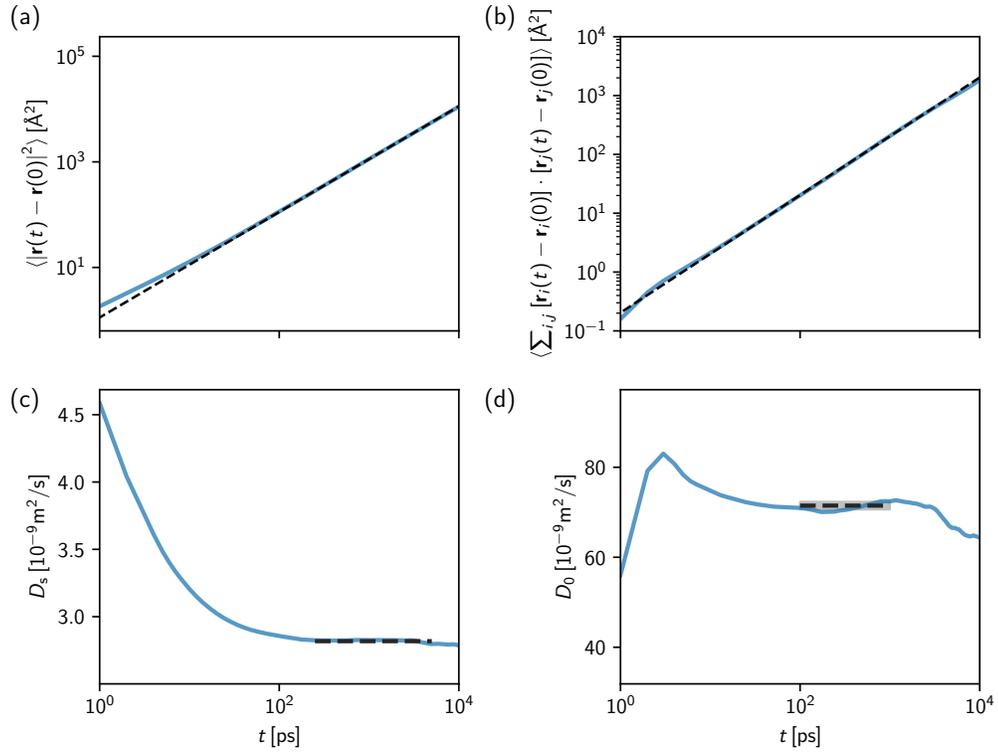

FIG. S7. **Analysis of the mean-squared displacement to obtain self and collective diffusion coefficients.** Data is exemplarily shown for the pore with modulus $E_{zz} = 2$ GPa and height $H = 0.9$ nm. Shown in (a) and (b) are the mean-squared displacements (MSD) $\langle |\mathbf{r}(t) - \mathbf{r}(0)|^2 \rangle$ and $\langle \sum_{i,j} [\mathbf{r}_i(t) - \mathbf{r}_i(t)] \cdot [\mathbf{r}_j(t) - \mathbf{r}_j(t)] \rangle$, respectively, that correspond to the self and collective diffusion. Instead of determining appropriate fitting ranges from this data, we show in (c) and (d) the corresponding MSD divided by ($2dt$) yielding directly the diffusion coefficient in the limit $t \to \infty$. The fitting result is shown as dashed horizontal line, the length of which indicates the fitting ranges (between 250-5000 ps for $D_s$ and 100-1000 ps for $D_0$, respectively). Dashed lines in (a) and (b) denote the MSD according to the fitted diffusion constant.

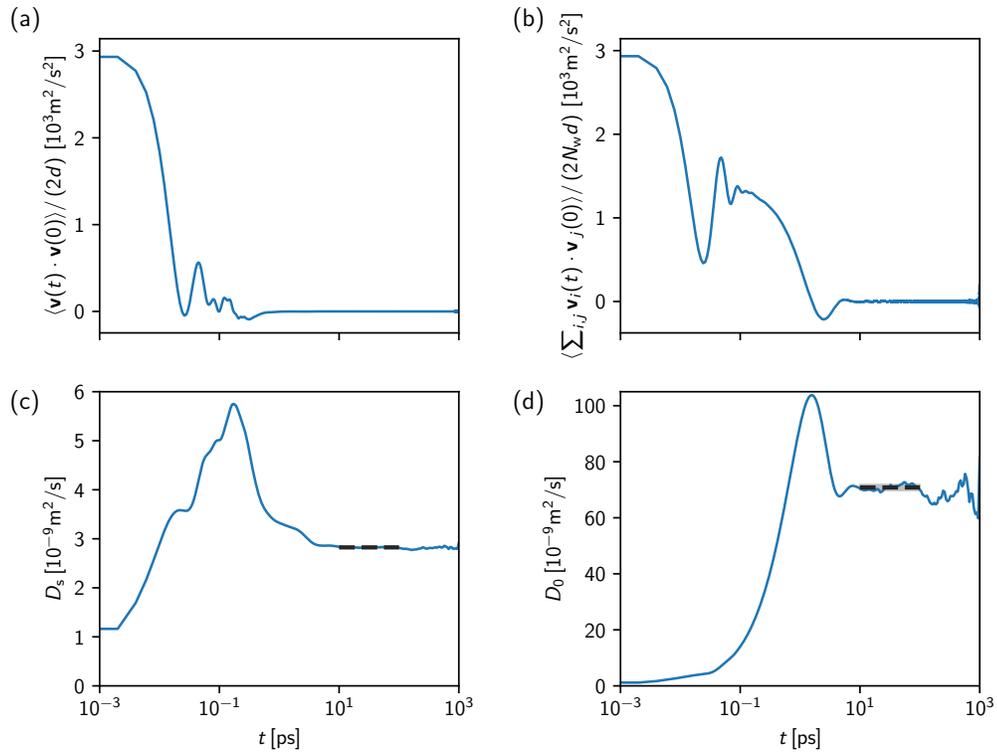

FIG. S8. **Analysis of the velocity auto-correlation function to obtain self and collective diffusion coefficients.** Data is exemplarily shown for the pore with modulus $E_{zz} = 2\,\mathrm{GPa}$ and height $H = 0.9\,\mathrm{nm}$ (as in Fig. S7). Shown in (a) and (b) are the velocity auto-correlation functions for the self and collective diffusivities, respectively. (c) and (d) denote the corresponding values of the integrals $D_0 = 1/(2N_{\mathrm{w}}dt)\langle\sum_{i,j}[\mathbf{r}_i(t) - \mathbf{r}_i(0)] \cdot [\mathbf{r}_j(t) - \mathbf{r}_j(0)]\rangle$ and $D_{\mathrm{s}} = 1/(2dt)\langle|\mathbf{r}(t) - \mathbf{r}(0)|^2\rangle$, which converge to the desired values in limit $t \to \infty$. Due to limited sampling of the auto-correlation function noise appears in the integrals, we this take the mean between 10 and 100 ps, indicated as dashed lines, to calculate $D_{\mathrm{s}}$ and $D_0$.

## II. CONSTRUCTION OF THE SLIT PORE

We construct a chemically realistic, disordered material for the slit pore based on a 25 Å³ sample of mineral-free shungite (PY02), a rather exceptional kerogen type from Russia, that was obtained using a molecular dynamics-hybrid reverse Monte Carlo approach by Bousige and co-workers [1]. This material only has tiny ratios of oxygen and hydrogen atoms compared to carbon atoms and thus is expected to be rather hydrophobic, as desired in our study. We chose a high density of 1.4 g/cm³ to obtain a nearly water impermeable material.

For this cubic sample we cut out a slice of 10 Å thickness and estimate the average charge on the carbon, hydrogen and oxygen atoms from a 100 ps simulation at $T = 300$ K using the Reax.ff reactive forcefield with charge equilibration [2, 3] and a timestep of 0.5 fs. The sample is then duplicated, rotated by 90° around the $z$-axis to avoid symmetry and by 180° around the $x$-axis to cancel $z$-component of dipole moment and then translated by $H$ in $z$-direction to create a slab system. To allow for pore fluctuations without interactions with the periodic images in $z$, a vacuum layer of 40 Å is added. In order to account for this already existing vacuum layer when computing the electrostatic interactions with the Yeh-Berkowitz correction,[4] the slab correction of LAMMPS is used together with another virtual vacuum layer of 1.5 times the box length, resulting in at least 3 times the box length of the vacuum thickness.

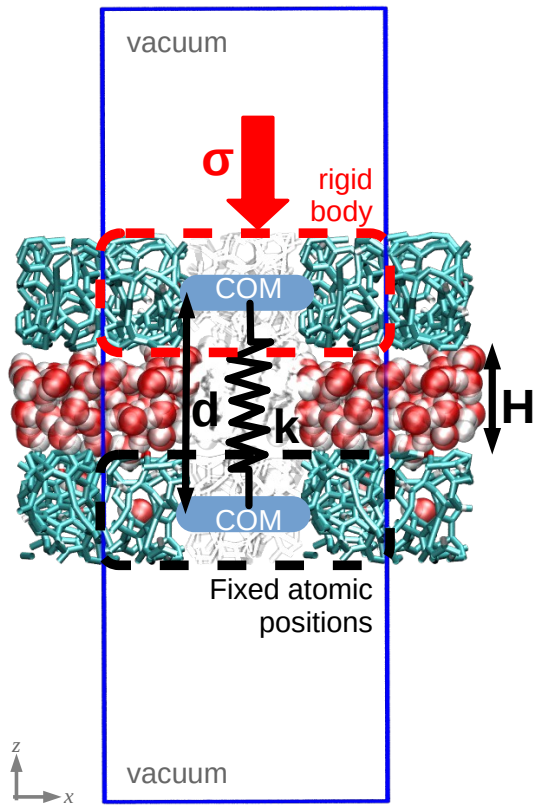

FIG. S9. **Setup of the simulations.** The constructed slit pore is periodic in the $xy$-plane and separated by a vacuum layer of 4 nm. A Hookean spring of equilibrium length $d$ is acting between the center of mass of the two surfaces resulting in the surface separation $H$. A stress $\sigma$ is applied on the fluctuating surface via a homogeneous force density $f_z$ in $z$-direction.

As explained in the main text, positions of the lower surface (in $z$-direction) are fixed, whereas the upper surface is allowed to fluctuate. A stress $\sigma_{zz}$ is applied via a homogeneous force density $f_z = \sigma_{zz} L_x L_y / N_{\text{wall}}$, where $L_x = L_y = 2.5$ nm are the lateral dimensions of the simulation box and $N_{\text{wall}} = 502$ is the number of atoms in one surface. Equations of motion are integrated in the thermalized $N_{\text{water}} V T$ ensemble for water molecules using the Nosé-Hoover thermostat at temperature $T = 300$ K, whereas the fluctuating rigid surface is thermalized only via energy exchange with the water molecules in the $N_{\text{wall}} V E$ ensemble at fluctuating energy $E$.

## III. BULK WATER EQUATION OF STATE

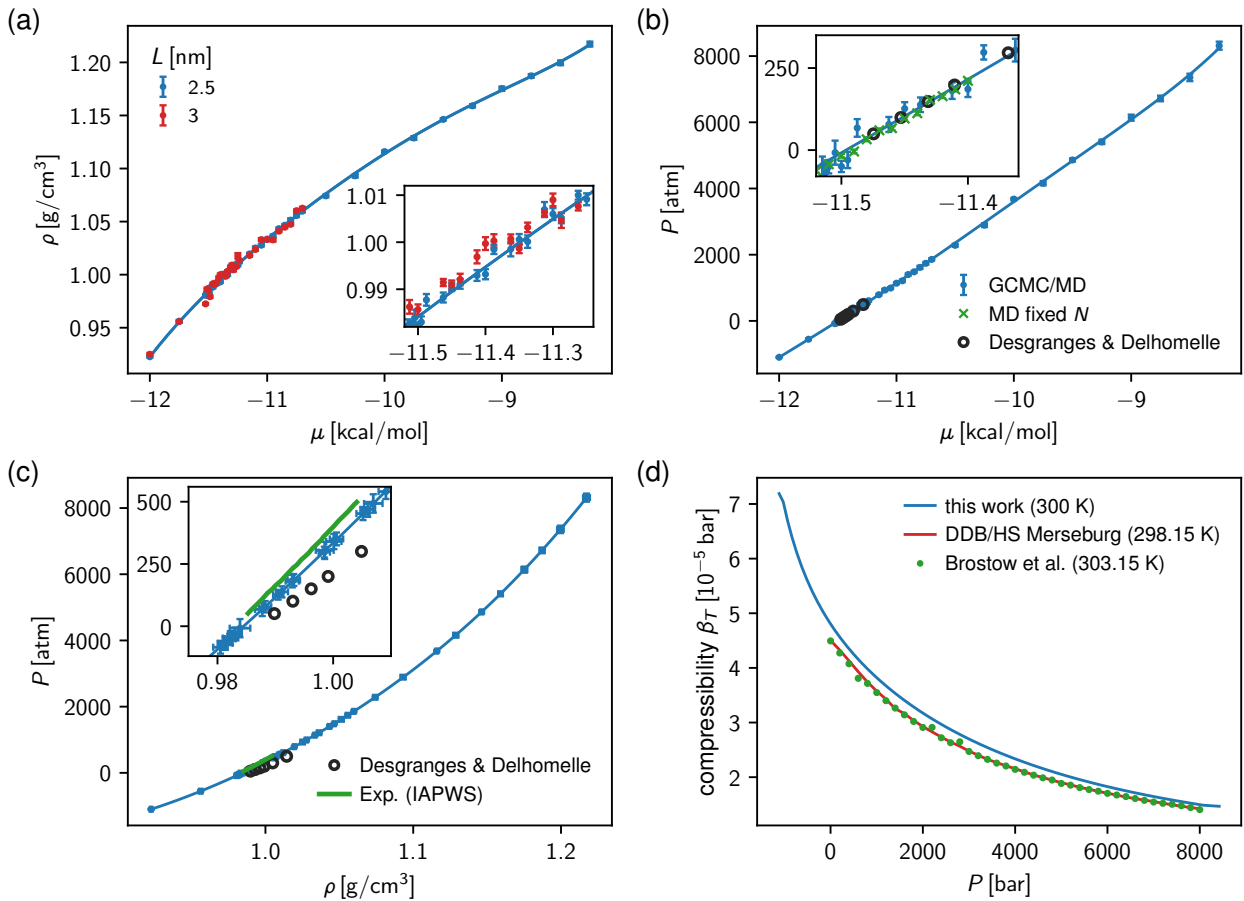

FIG. S10. **GCMC/MD simulations of bulk water in the liquid state.** (a) Density for two box lengths $L$ indicated in the legend and polynomial fit to the data for $L = 2.5$ nm. (b) Pressure obtained from the GCMC/MD runs (blue data) and corresponding polynomial fit. Green data show simulation results obtained from MD simulations at fixed particle number $N = \rho L^3$ for the density determined in (a) and $L = 2.5$ nm. Empty black circles denote simulation data from Ref.[5] obtained in a 2 nm box and using analytical long-range correction for the truncated Lennard-Jones interaction. (c) shows the resulting equation of state together with simulations from Ref.[5] and experimental values taken from the IAPWS tables, Ref.[6]. Isothermal compressibility of SPC/E water using our simulation parameters (blue line) and experimental results from the Dortmund Data Bank, 2022, www.ddbst.com (red line) and Ref.[7] recorded at temperatures $T = 298.15$ K and $T = 303.15$ K respectively.

GCMC/MD simulations have been performed for bulk water in a cubic box of box length $L = 2.5$ nm at different values of the chemical potential $\mu$. As for the slab system discussed in the main text every 1 ps a GCMC run of $2 \times 10^4$ steps was performed starting from a configuration at typical bulk water density and the total sampling was conducted for $10^6$ GCMC steps to obtain convergence with typically 5% of the data being discarded for equilibration; error estimates are obtained by block-averaging the simulation data into 20 independent samples. For the bulk simulations of the SPC/E water kept rigid via the SHAKE algorithm[8] we employ a MD timestep of 2 fs and the Nosé-Hoover thermostat is set to a characteristic damping time of 0.2 ps.[9]

Figure S10(a) shows the resulting density of liquid bulk water for two box lengths $L \in \{2.5, 3\}$ nm to exclude possible finite size effects. The solid line is obtained from a fifth-order polynomial fit to the simulation data. Comparison of the two box lengths reveals no finite size effects. Figure S10(b) shows the corresponding pressure determined from the GCMC/MD run (blue data). Since the fluctuating particle number and the corresponding thermal equilibration of all degrees of freedom within the 1 ps MD run might result in insufficient phase-space sampling we also performed MD simulations at fixed water particle number $N_{\text{water}}$ determined from the polynomial fit shown in Fig. S10(a).

The resulting pressures shown as green crosses in Fig. S10(b) perfectly agree with the GCMC/MD data and the corresponding polynomial fit (blue line). We also include in Fig. S10(b) simulation data by Desgranges and Dellhomme obtained via Wang-Landau sampling in a 2 nm cubic simulation box using analytic long-range correction for the truncated Lennard-Jones potential (empty black circles) which perfectly matches our simulation results. From the polynomial fit of the $P - \mu$ relation shown in the inset of Fig. S10(b) we choose the value of the chemical potential employed in the main text, $\mu = -11.4$ kcal/mol which corresponds to target intrusion pressure introduced in the main text, $P_0 = 215$ atm.

In Fig. S10(c) we show the water equation of state resulting from our simulations (blue data and blue line denoting a polynomial fit) and experimental results taken from the IAPWS tables,[6] revealing excellent agreement. Interestingly, employing the analytical dispersion correction as in the work by Desgranges and Dellhomme leads to worse results compared to the experimental values for SPC/E water. The polynamial fit in Fig. S10(c) allows to calculate the isothermal compressibility

$$\beta_T = -\frac{1}{v}\frac{\partial v}{\partial P},$$ (S1)

where $v = m_{\text{water}}/\rho$ is the volume of a water molecule and $m_{\text{water}}$ is its mass. Comparison of our simulation data at $T = 300$ K [blue line in Fig. S10(d)] with experimental results for $\beta_T$ at $T = 298.15$ K and 303.15 K, respectively, reveals excellent and near-quantitative agreement for the thermodynamic-mechanical properties and thus justifying the choice of the SPC/E model for this work.

## IV. DIFFUSION COEFFICIENT IN BULK WATER

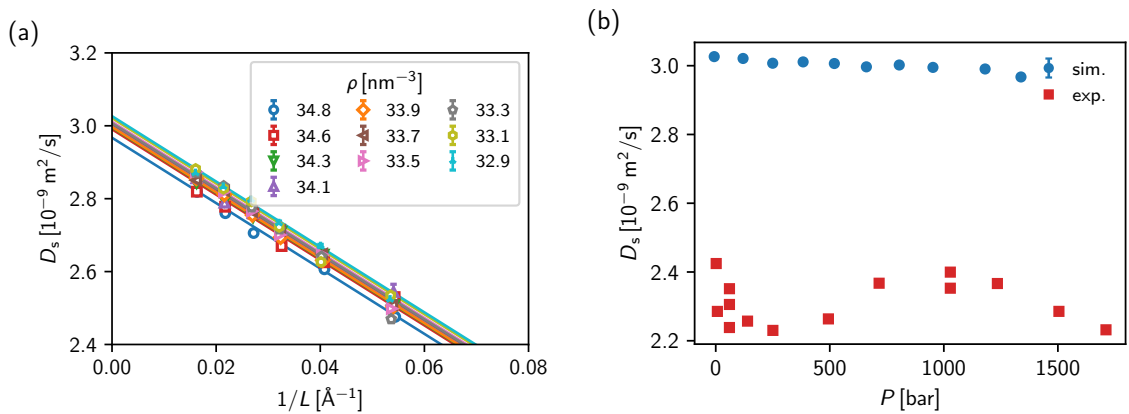

FIG. S11. **Self-diffusion coefficient in bulk water.** (a) Rigorous finite size scaling analysis of $D_S$ vs. the inverse cubic box size $L$. (b) Extracted self-diffusion coefficient $D_S$ at infinite system size as a function of the pressure $P$ in the liquid phase. Simulations denote results from this work extracted in (a), experimental data are taken from Krynicki et al..[10]

To correct for the finite size of our cubic simulation box of length $l$, we follow the finite-size scaling approach by Yeh and Hummer to obtain the self-diffusion coefficient in an infinite nonperiodic system,[11]

$$D_s^{(\text{PBC})} = D_s^{(\text{bulk})} - \frac{k_B T \xi}{\pi \eta_b L}.$$ (S2)

Here, $D_s^{(\text{bulk})}$ is the self-diffusion coefficient in the non-periodic, infinite-size bulk liquid phase, whereas $D_{\text{PBC}}$ is the corresponding value obtained in a finite, periodic simulation box. The thermal energy is $k_B T$ and the shear viscosity is $\eta_b$.

Figure S11(a) shows the self-diffusion coefficient of bulk water determined from the mean square displacement for different densities in a constant volume and temperature ($NVT$) simulations, where the side length $L$ is varied. The solid lines in Fig. S11(a) denote fits of the simulation data to Eq. (S2), revealing excellent agreement with the expected scaling $L^{-1}$. The corresponding extrapolated values $D_s^{(\text{bulk})}$ are shown in Fig. S11(b) as blue symbols and are practically independent of the pressure. During the fitting procedure, we fixed the value $\eta_b = 0.695$ mPa s

independently of the density.[12] If we treat the $\eta_b$ as free parameter during fitting, the resulting bulk viscosity independent of pressure within the fitting errors and independent of pressure in the considered range, $\eta_b^{(fit)} = 0.696 \pm 0.017\,\mathrm{mPa\,s}$, well in line with the literature,[12] and reflecting again the fact that water is incompressible in this pressure range. We also include in Fig. S11(b) the corresponding pressure-dependent water self-diffusion coefficient measure using the proton spin echo method.[10] Whereas the employed SPC/E water model over-estimates the water self-diffusion by about 30%, the fact that $D_s$ is independent of pressure is well reflected within the experimental values.

---

# TOC Graphic

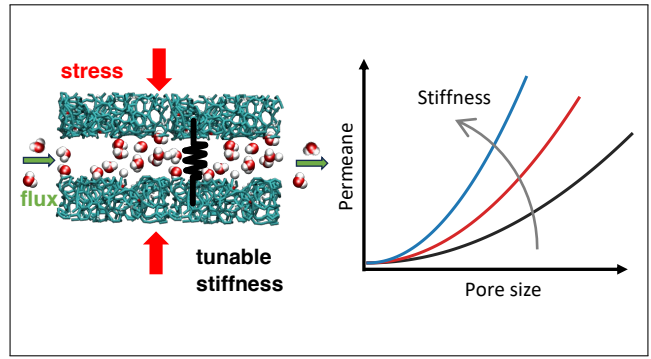


\fi

\begin{tocentry}
	\includegraphics{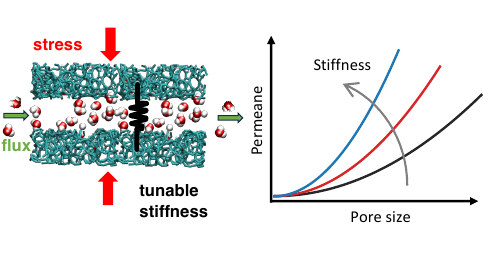}
\end{tocentry}

\end{document}